%% file: rieutord_valdettaro.tex
\documentclass{jfm}
\usepackage{subfigure}
\usepackage[dvips]{graphicx}
\usepackage[latin1]{inputenc}
\usepackage{psfrag}
\usepackage{rotating}
\usepackage{natbib}
\usepackage{bm}

\title{
Viscous dissipation by tidally forced inertial modes in a rotating
spherical shell }

\author[M. Rieutord and L. Valdettaro]{
M. RIEUTORD$^{1}$ and L. VALDETTARO$^{2}$ 
}

\affiliation{
$^1$Laboratoire d'Astrophysique de Toulouse - Tarbes, CNRS et
Université de Toulouse, 14 avenue E. Belin, 31400 Toulouse,
France\\
$^2$MOX - Dipartimento di Matematica, Politecnico di Milano, Piazza L.
da Vinci, 32, 20133 Milano, Italy
}

\pubyear{2009}
\volume{000}
\pagerange{000--000}
\date{\today}
\input com  % Attention commandes locales a causes d'interferences avec jfm
\def\dd{{\rm d}}

\begin{document}
\bibliographystyle{jfm}
\maketitle
%%%%%%%%%%%%%%%
%
%  a b s t r a c t
%

\begin{abstract}
We investigate the properties of forced inertial modes of a rotating
fluid inside a spherical shell. Our forcing is tidal like, but its main
property is that it is on the large scales. By numerically solving the linear
equations of this problem, including viscosity, we first confirm some
analytical results obtained on a two-dimensional model by \cite{O05}; some
additional properties of this model are uncovered like the existence of
narrow resonances associated with periodic orbits of characteristics. We
also note that as the frequency of the forcing varies, the dissipation
varies drastically if the Ekman number $E$ is low (as is usually the case).
We then investigate the three-dimensional case and compare the results
to the foregoing model. The three-dimensional solutions show, like their
2D counterpart, a spiky dissipation curve when the frequency of the
forcing is varied; they also display small frequency intervals where the
viscous dissipation is independent of viscosity. However, we show that the
response of the fluid in these frequency intervals is crucially dominated
by the shear layer that is emitted at the critical latitude on the inner
sphere. The asymptotic regime, where the dissipation is independent of
the viscosity, is reached when an attractor has been excited by this
shear layer. This property is not shared by the two-dimensional model
where shear layers around attractors are independent of those emitted at
the critical latitude.  Finally, resonances of the three-dimensional
model correspond to some selected least-damped eigenmodes. Unlike
their two-dimensional counter parts these modes are not associated with
simple attractors; instead, they show up in frequency intervals with
weakly contracting webs of characteristics. Besides, we show that the
inner core is negligible when its relative radius is less than the
critical value $0.4E^{1/5}$. For these spherical shells, the full
sphere solutions give a good approximation of the flows.
\end{abstract}

\section{Introduction}

\subsection{The background}

The question of how close binary stars born on eccentric orbits reach
a circular trajectory with synchronous orbital and spin rotation is by far
not fully resolved. The process of circularization of the orbits and
synchronization of rotations is thought to be the result of the tidal
interaction. Tides act in two ways: they first create a tidal bulge
which generates a torque leading to angular momentum exchange between the
orbital motion and the spinning stars.  Second, they dissipate energy
through the induced fluid flows. This latter effect is important as it
controls the time scale of the whole evolution process.

Thus, much work has been devoted to evaluate the efficiency of the
dissipative processes, which can be triggered by tides.  The first
process that has been investigated is the viscous damping of the
equilibrium tide; this is the tide induced by the mere deformation of
equipotentials; the fluid is assumed ``attached" to these surfaces
and the time dependence of the tidal distortion makes the flow (see
\citealt{zahn66,zahn77,zahn92,zahn08}). However, this process is efficient if
the viscosity is high enough, a condition which is met only by low-mass
stars. Indeed, the convective envelope of these stars provides a strong
damping through their turbulent viscosity. However, stars of mass
larger than 1.8 solar mass have no convection in their outer layers
(only in the central part).  Their envelope is stably stratified. But
observations do show that circularization and synchronization are also
effective for these stars when they are close enough (\citealt{gmm84}, but
see also the recent review of observational facts on tidal
interactions between stars or between stars and planets by
\citealt{Mazeh08}). The
present understanding of these results is based on the idea that the
radiative envelope of these stars are efficiently damping the gravity
waves excited by the tides. This mechanism was first investigated by
\cite{zahn75} using a simplified approach with an asymptotic description
of gravity modes in a non-rotating fluid. However, stars in a binary
system are rotating, sometimes quite rapidly. Rotation thus appeared as
an unavoidable feature. Following work of \cite{rocca87,rocca89} included
therefore the Coriolis acceleration but as a first order perturbation.

More recently, \cite{WS99a,WS99b,WS01} investigated numerically the
synchronization process of massive stars, fully taking into account
the Coriolis force in the flow. They thus discovered the phenomenon of
``resonance locking" by which the tidal forcing excites resonantly two
rotationally modified gravity modes (hereafter called gravito-inertial
modes), which therefore strongly dissipate energy; the  locking occurs
because one mode tends to spin the star up while the other tends to spin
it down. The low frequencies of the resonant gravito-inertial modes,
however, is synonymous of short-wavelength modes that are not well
suited to the foregoing two-dimensional numerical calculations. This
difficulty motivated \cite{SW02} to further investigate the problem
with the so-called Traditional Approximation which allows a separation
of the variables. However, as discussed by \cite{Gerkema_etal08},
this simplification eliminates a crucial feature of these modes,
namely their internal shear layers. Indeed, \cite{DRV99} showed
that gravito-inertial modes are singular in the limit of vanishing
diffusivities. This property is actually shared by gravity modes and
inertial modes in any configuration where variables cannot be separated
\cite[see][]{RGV00}. The singularities come from the ill-posed nature
of the mathematical problem which is of hyperbolic or mixed type with
boundary conditions. At non-zero but low diffusivities, dominant
singularities show up as modes confined in shear layers following
attractors of characteristics \cite[][]{RV97,RGV01}. As far as dissipation
is concerned, such modes behave very differently compared to the regular
ones. Astrophysical situations being characterised by large ratios
between integral and dissipation scales, it is crucial to understand
the asymptotic properties of resonant-mode dissipation at vanishing
viscosities and thermal diffusion.

Besides the problem of tidal interactions between stars, which is
quite old, the recent discovery of many planetary systems harbouring
Jupiter-like planets on orbits very close to the central star also
motivates a new examination of the tidal interaction. The novelty with
planets, even of the size of Jupiter, is that they likely contain a
rocky core in their central part. Even if this core is not a solid
body, the transition with the surrounding envelope is likely sharp
\cite[][]{GL09}. This makes the fluid domain like a spherical shell. In
stars this domain is delineated by the \BVF\ variations and may be more
complex \cite[e.g.][]{DR00}.

Beyond the astrophysical problem described above, the case of resonant
inertial modes is also of interest in Earth sciences in relation to
the elliptic instability \cite[e.g.][]{LaLL05}, the rotating precessing
flows \cite[e.g.][]{HK95} or for the understanding of the dynamics of the
ocean or of the atmosphere \cite[e.g.][]{Maas01,MH07}. Furthermore, it
has also been investigated in the context of engineering applications
such as the oscillations of fuel tanks of spinning spacecrafts
\cite[see][]{Mana96}. Finally, let us mention that non-axisymmetric
inertial modes also appear as the growing perturbations of a rotating
fluid destabilized by thermal convection \cite[see][]{Z94,Z95}.

As may be guessed, the full astrophysical problem is very involved
and some simplifications are in order if one wishes to decipher the
mechanisms controlling the asymptotic laws of dissipation at small
diffusivities. Thus, following the work of \cite{O05}, we first
investigate the case of forced singular modes in a slender toroidal
shell, which is a two-dimensional approximation for a spherical shell
\cite[e.g.][]{RVG02}. Thus doing, we can recover and extend the results of
\cite{O05}. We proceed then by focusing on the three-dimensional problem
of the spherical shell and extend the previous work of \cite{R91}
and \cite{tilg99b}. We then discuss and compare the results of 2D and
3D models. Some conclusions end the paper.

\subsection{The model}

We consider a viscous fluid inside a spherical shell that mimics a
stellar or planetary envelope, submitted to the tides of an orbiting mass point. This
tidal forcing may be condensed in the tidal potential which we write as:

\[ \Phi_T= \Phi_{\rm ax} r^2 P_2(\cth)\cos\omega_o t + \Phi_{\rm nx} r^2
P_2^2(\cth)\cos(2\omega_o t -2\varphi) \]
following \cite{zahn77}. In this expression, $\omega_o$ is the orbital
angular velocity of the point mass, $(r,\theta,\varphi)$ are the spherical
coordinates whose origin is at the centre of the body under consideration;
$P_2$ and $P_2^2$ are Legendre polynomials, while $\Phi_{\rm ax}$
and $\Phi_{\rm nx}$ are the amplitudes. On Earth $\Phi_{\rm nx}$ is
the amplitude of the well-known semi-diurnal tide.\linebreak $\Phi_{\rm ax}r^2
P_2(\cth)\cos\omega_o t$ is the leading term coming from the eccentricity
of the orbit. Although $\Phi_{\rm ax}$ is usually smaller than $\Phi_{\rm
nx}$, we shall discard $\Phi_{\rm nx}$ and only consider the first
term of $\Phi_T$. This is justified by the fact that the properties of
non-axisymmetric inertial modes in a spherical shell are the same as
those of the axisymmetric ones as far as singularities are concerned
\cite[][]{RGV01}. Thus all the results derived below for an axisymmetric
forcing can be applied, {\it mutatis mutandis}, to a non-axisymmetric
forcing. In another problem like the resonant interaction of inertial
modes, which leads to the elliptic instability \cite[][]{kers02}, the
non-axisymmetric terms would be essential of course.

The fluid inside the spherical shell is rotating at the spin angular
velocity $\Omega$ of the star. Thus, in a frame co-rotating with the fluid,
the tidal forcing is 

\[ \Phi_T= \Phi_{\rm ax} r^2 P_2(\cth)\cos[(\omega_o-\Omega) t]
\]
The first response of a star to the tidal potential is the so-called
equilibrium tide, which describes the distortion of the equipotentials. This
induces a velocity field $\vv_e$, which is derived from the time-evolution
of the equipotentials. Following \cite{zahn66}, this induces a radial
flow of the form:

\[ \vV_e = \calA r^2 P_2(\cth)\sin\omega t\er
\]
where $\omega=\omega_o-\Omega$.
As shown by \cite{O05}, this flow forces a dynamical response of
the star through the body force

\[ \vf = -(i\omega \vV_e + 2\vO\times\vV_e)\]

We are interested in the dynamical response of the
fluid. As the general realistic case is much involved, we
reduce it to the study of the response of an incompressible viscous
fluid inside a spherical shell. Although much simplified, this model
retains the essential feature of the low-frequency stellar modes, namely
their singularities associated with attractors of characteristics
\cite[e.g.][]{DR00}.

Assuming that the fluid response remains of small amplitude, we need to
solve the linearised equations governing forced periodic perturbations
of a viscous rotating fluid with constant density. When the length scale
is the outer radius of the shell $R$, the time scale is
$(2\Omega)^{-1}$, the equations of the non-dimensional pressure ($p$)
and velocity perturbation ($\vu$) may be written:

\greq
i\omega\vu + \ez\times\vu  = -\na p + E\Delta\vu +\vf\\
\\
\Div\vu = 0
\egreqn{eqmotion}
where $E=\nu/2\Omega R^2$ is the Ekman number and $\nu$ is the kinematic
viscosity. The non-dimensional force reads

\beq \vf = -V_e(r,\theta)(i\omega\er+ \sth\ephi) \eeqn{force}
where $V_e(r,\theta)=\calA r^2 P_2(\cth)$. We complete these
equations with stress-free boundary conditions. The use of
stress-free boundary conditions may be surprising in the planetary
case as the interface between the fluid layer and the core is a solid
boundary. However, because of the nature of the fluid flows that are
restricted to internal shear layers, this boundary condition has little
effect on the solution as was actually found by \cite{FH98}. We give in
appendix~B the scaling arguments that lead to this result.

Even thus simplified, our problem remains challenging. A further
simplification, actually used many times \cite[][]{ML95,RVG02,O05},
consists in reducing the problem to two dimensions. In such a case, the
spherical shell turns into a (cored) slender torus \cite[see
figure~\ref{slender_torus} and][]{RVG02}, but the
eigenvalue problem as well as the forced one (far from resonances) can
be solved analytically \cite[][]{RVG02,O05}. Although this latter
simplification seems quite drastic, it may be shown that it is
nevertheless relevant to equatorial regions of a thin spherical shell
\cite[e.g.][]{stewar71,RVG02}. To bridge the gap with the work
of \cite{O05}, we therefore first consider this two-dimensional model and
then move on to the three-dimensional one.

\section{Resonances of axisymmetric modes in the slender torus}

\subsection{Mathematical formulation}

Using cylindrical coordinates $(s,\varphi,z)$, we first introduce
a meridional stream function $\psi$ so that mass conservation is
automatically insured. Hence, we set

\[u_s = -\dz{\psi}, \qquad u_\varphi = u, \qquad u_z = \frac{1}{s}\ds{s\psi}\; .
\]
The momentum equation leads to 

\greq
E\Delta'\Delta'\psi-\partial_zu -i\omega\Delta'\psi = C \\
E\Delta'u + \partial_z\psi - i\omega u = -f
\egreqn{eqmou}
where

\[ \Delta' = \Delta - 1/s^2 = \ds{}\lp\frac{1}{s}\ds{}s\rp+ \ddz{}\]
and

\[
f=f_\varphi =  -\sth V_e(r,\theta)\qquad C=\dz{f_s}-\ds{f_z}\]

\[ f_s=-i\omega\sth V_e(r,\theta)\qquad f_z=-i\omega\cth V_e(r,\theta)
\]
The torque density $C$ may be further reduced into

\[ C = \frac{i\omega}{r}\dtheta{V_e}\]

System \eq{eqmou} is completed by stress-free boundary conditions which
demand that

\[ \psi=\dr{}\lp\frac{1}{r}\dr{\psi}\rp=\dr{u}=0 \qquad {\rm at}\quad r=\eta,1\]
Here $\eta$ is the non-dimensional radius of the inner core.

\begin{figure}
\centerline{
\includegraphics[width=0.9\linewidth,clip=true]{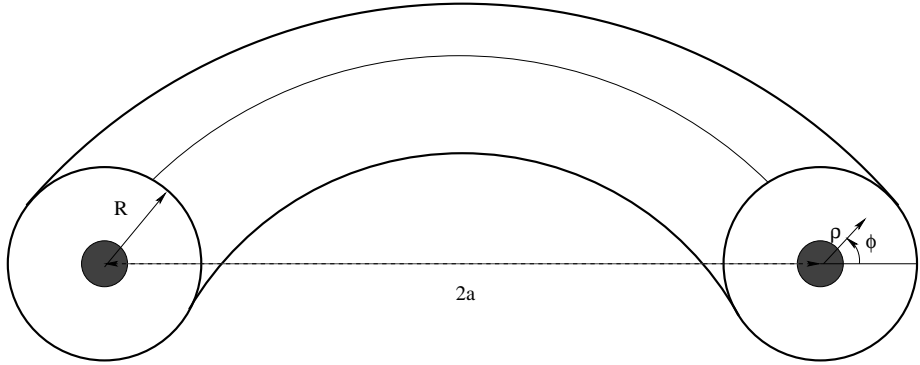}}
\caption[]{A schematic view of the slender cored torus: the principal
radius $a$ of the torus tends to infinity.}
\label{slender_torus}
\end{figure}

\subsection{Solving the equations}

Because of the shape of the boundaries, it is natural to use polar
coordinates $(\rho,\phi)$ in a meridional section of the torus (see the
schematic view in figure~\ref{slender_torus}). These coordinates are
related to the foregoing cylindrical coordinates by

\[ s=\rho\cos\phi, \qquad z=\rho\sin\phi\]
They are also related to the original spherical coordinates by $r=\rho$ and
$\theta=\frac{\pi}{2}-\phi$.

The two-dimensional approximation is essentially summarised into the
neglect of the curvature terms so that, for instance, $\psi/s \ll
\partial\psi/\partial s$; hence,

\begin{eqnarray*}
 u_\rho&=&u_s\cphi  + u_z\sphi = -\frac{1}{\rho}\dphi{\psi} \\
 u_\phi&=&-u_s\sphi + u_z\cphi = \drho{\psi} 
\end{eqnarray*}
To obtain a numerical solution of the equations \eq{eqmou}
(simplified with the 2D-approximation), we first use a Fourier
decomposition, namely

\[ (\psi,u,f,C)(\rho,\phi) =\sum_n (\psi_n, -iV_n, f_n, C_n)(\rho)e^{in\phi}, \]
and find the set of ordinary differential equations controlling the shape
of the radial functions $\psi_n(\rho)$ and $V_n(\rho)$. They are

\greq
E\Delta_n V_n + \frac{\psi'_{n-1}-\psi'_{n+1}}{2} - \frac{(n-1)\psi_{n-1}+
(n+1)\psi_{n+1}}{2\rho} -i\omega V_n = -if_n  \\
\\
E\Delta_n\Delta_n\psi_n + \frac{V'_{n-1}-V'_{n+1}}{2} -\frac{(n-1)V_{n-1}+
(n+1)V_{n+1}}{2\rho} - i\omega\Delta_n\psi_n = C_n
\egreqn{setode}
where

\[ \Delta_n = \ddrho{}+\frac{1}{\rho}\drho{}-\frac{n^2}{\rho^2}\]
With this formulation, the stress-free boundary conditions read:

\[ \psi_n=\drhorho{\psi_n}-\frac{1}{\rho}\drho{\psi_n}=\drho{V_n}=0\]

For the tidal forcing at hands, we find that

\greq
f(\rho,\phi) = \demi\rho^2(\cphi-2\sin^2\phi\cphi)=\frac{\rho^2}{8}(\cphi+3\cos3\phi)\\
C(\rho,\phi) = -3i\omega\rho\sphi\cphi
\egreqn{nat_force}
so that

\[ f_n=f_{-n}=\frac{\rho^2}{16}(\delta_{n,1}+3\delta_{n,3}), \qquad C_n=-C_{-n}=
-\frac{3}{4}\omega\rho\delta_{n,2} \]
The two-dimensional set-up restricts the number of excited attractors
compared to the three-dimensional one \cite[see][and below]{RVG02}. Thus,
although this forcing is derived from the true tidal force, we complete
our view of the solutions by the use of the following forcing

\[ f(\rho,\phi) = 2\cos2\phi, \qquad C=0;\]
thus

\[ f_n=f_{-n}= \delta_{n,2}, \qquad C_n =0\]

\subsection{Symmetries}

In modelling the tidal interaction, it is commonly assumed that spin and
orbital angular momentum vectors are parallel. With this assumption,
the tidal flow is symmetric with respect to the equatorial and orbital
planes. In this case, the tidal force is such that $f_n=f_{-n}$ and
$C_n=-C_{-n}$ as noted above.

This symmetry allows us to solve the set of equations \eq{setode} solely
for $n\geq0$. However, solutions in the torus verify a further
symmetry: they may be symmetric or antisymmetric with respect to the
transformation $\phi\tv\phi+\pi$.  Such a symmetry is specific to the
torus and does not exist in the sphere. As noticed by \cite{RVG02}, it
has a selection effect on attractors: some may exist in the spherical
shell but be not authorised in the torus. For this reason, some
attractors of the spherical shell cannot be studied in two dimensions
with the natural forcing \eq{nat_force}, but can be investigated with the second
forcing. Still some others cannot be studied in two dimensions
altogether.

The equatorial symmetry of the tidal forcing in combination with the
parity of its Fourier component, implies that only half of the Fourier
components are excited, namely

\[ V_1,V_3,\ldots, V_{2n+1},\ldots \qquad \psi_2,\ldots,\psi_{2n}\ldots\]
we also note that, since $V_{-1}=V_1$,

\[ \Delta\Delta\psi_0 - i\omega\Delta\psi_0=0\]
which means that $\psi_0$ is not excited and therefore vanishes. The
second forcing excites the other set of Fourier components, namely

\[ V_0, V_2,\ldots, V_{2n},\ldots \qquad
\psi_1,\ldots,\psi_{2n+1}\ldots\]

\subsection{Dissipation}

As discussed in the introduction, total viscous dissipation of the
fluid volume is the actual quantity to be evaluated when the solution is
known (this is the quantity which controls the secular evolution of the
orbit). As noted by \cite{O05}, dissipation may be evaluated in two ways:

\beq D=\frac{E}{2}\intvol|c|^2 dV =\intvol Re(\vu^*\cdot\vf) dV
\eeqn{eq:dissip}
where $[c]$ is the rate-of-strain tensor. We use both of these expressions to
evaluate the internal numerical precision of our results.

Let us mention that the components of the rate-of-strain tensor $[c]$ in
the $(\rho,\phi)$ coordinates are easily obtained if these
coordinates are completed by an axial one, let say $\zeta$, so that
$(\rho,\phi,\zeta)$ form a cylindrical system of coordinates (in
our problem, $\zeta$ is perpendicular to the meridional plane); we show
in appendix the equivalence of these expressions with the usual ones. In
these coordinates, the rate-of-strain tensor components are

\[ c_{\rho\rho} = c_{\phi\phi} =
2\lp\frac{1}{\rho}\drhodphi{\psi}-\frac{1}{\rho^2}\dphi{\psi}\rp, \qquad
c_{\rho\phi} = \ddrho{\psi}-\frac{1}{\rho}\drho{\psi}-
\frac{1}{\rho^2}\ddphi{\psi}\]
\[ c_{\rho\zeta} = \drho{u} , \qquad c_{\phi\zeta} = \drhophi{u}\]
and the volumic dissipation reads

\[ D= 2
(|c_{\rho\rho}|^2+|c_{\rho\zeta}|^2+|c_{\phi\zeta}|^2+|c_{\rho\phi}|^2)\]

\begin{figure}
\centerline{
\includegraphics[width=\linewidth,clip=true]{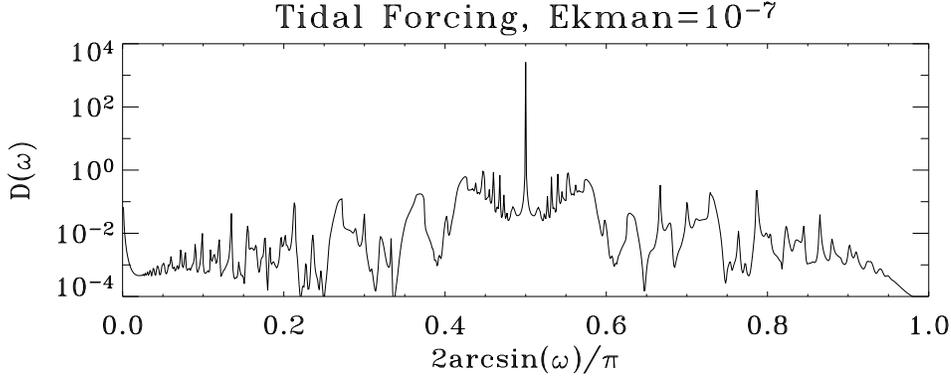}}
\caption[]{Scan of the dissipation as a function of the critical
latitude angle at $E= 10^{-7}$. Resolution is Nr=180, N$_\phi$=400.}
\label{scan_tf}
\end{figure}

\begin{figure}
\centerline{
\includegraphics[width=\linewidth,clip=true]{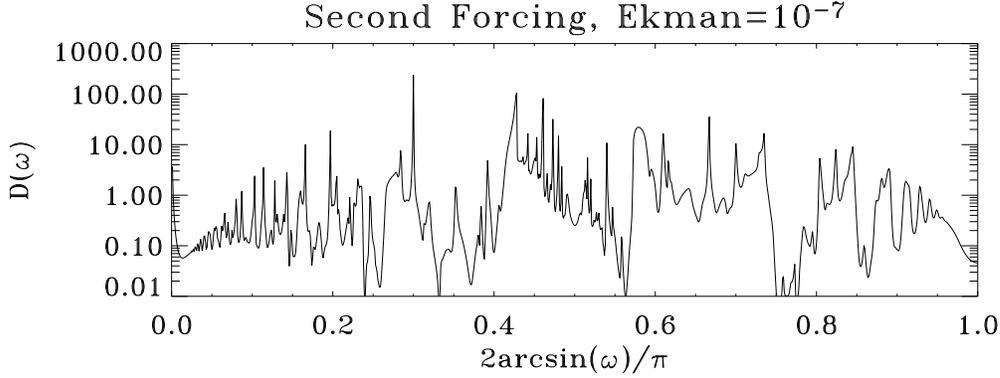}}
\caption[]{Same as in figure~\ref{scan_tf}, but for the second forcing.}
\label{scan_sym}
\end{figure}

\subsection{Results}

For comparison with previous work, in all the numerical applications,
we set the ratio of the inner radius to the outer one $\eta$ to 0.35
which is the value of the Earth's liquid core.

\subsubsection{Overall properties}

As shown by \cite{WS99b}, during the evolution of the orbits, the tidal
forcing frequency scans the whole inertial band and resonances there
play a major part. Let us point out that in the text below, we shall
understand the term ``resonance" as the local maximum (in frequency) of
the viscous dissipation. This definition is appropriate for our problem
but slightly different from the usual one which refers to the amplitude
of the flow. Resonances are usually understood as the signature of the
excitation of eigenmodes. As will be clear later, this classical view
is not always appropriate here because of the ill-posed nature of the inviscid
problem. Thus our wording ``resonance" should be understood in a rather
loose way.

With the tidal forcing in mind, we simulate a scan of the inertial
band. We represent the viscous dissipation as a function of the
frequency, or, more appropriately, as a function of the critical latitude
$\vartheta=\arcsin(\omega)$. As shown by figures~\ref{scan_tf} and
\ref{scan_sym}, the curve $D(\omega)$ is very spiky and almost symmetric
with respect to $\pi/4$. This underlying symmetry is a consequence of an
invariance of Poincaré equation in two-dimensions. Indeed, setting $E=0$
in \eq{eqmou} and rewriting the equation for $\psi$, we find

\[ \dds{\psi} -\lp\frac{1-\omega^2}{\omega^2}\rp\ddz{\psi} =
F(s,z,\omega)\]
in the two-dimensional limit. If $F=0$ we easily see that this equation
is invariant with respect to the transformation $s\rightarrow z$,
$z\rightarrow s$ and $\omega\rightarrow \sqrt{1-\omega^2}$. Thus the place
of resonant frequencies is indeed symmetric with respect to $\pi/4$, if
the critical latitude is used as a variable. However, the final curves,
which are shown in figures~\ref{scan_tf} and \ref{scan_sym}, partially
lose their symmetry because the forcing does not verify the foregoing
invariance as well as the viscous terms (the second forcing breaks more
strongly this symmetry).

\begin{figure}
\centerline{ \includegraphics[width=0.4\linewidth,clip=true]{coupe.eps}
\includegraphics[width=0.5\linewidth,clip=true]{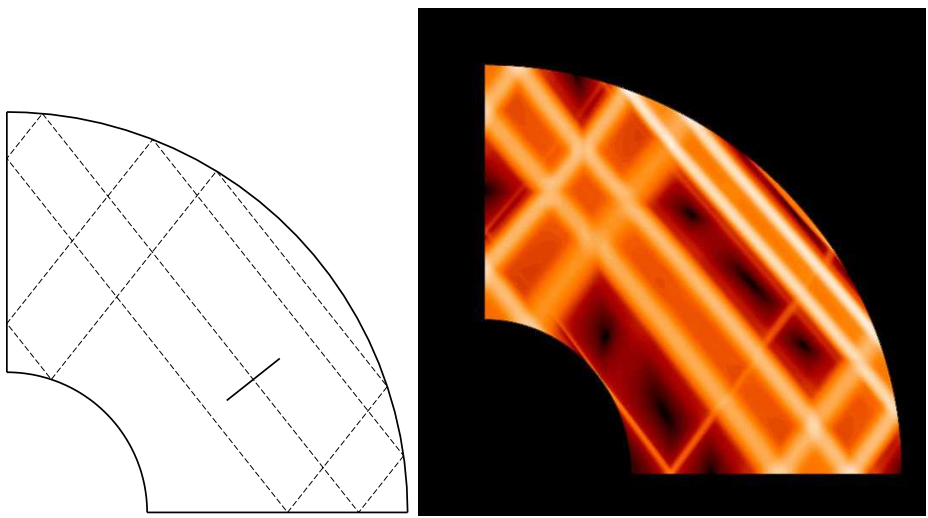} }
\caption[]{The attractor occupying the frequency band
[0.60922,0.62276]. Although prominent in the 3D problem, this attractor
cannot be excited by the natural tidal forcing; fortunately it can be
studied with the second forcing. The small segment indicates the place
where the velocity profiles of figure~\ref{profiles} have been taken. On
the right we show the kinetic energy distribution in a meridional plane
for a numerical solution computed at $\omega=0.621$ and $E=10^{-9}$.}

\label{attractor}
\end{figure}

\subsubsection{Attractors and associated
resonances}\label{section:attractors}

\begin{figure}
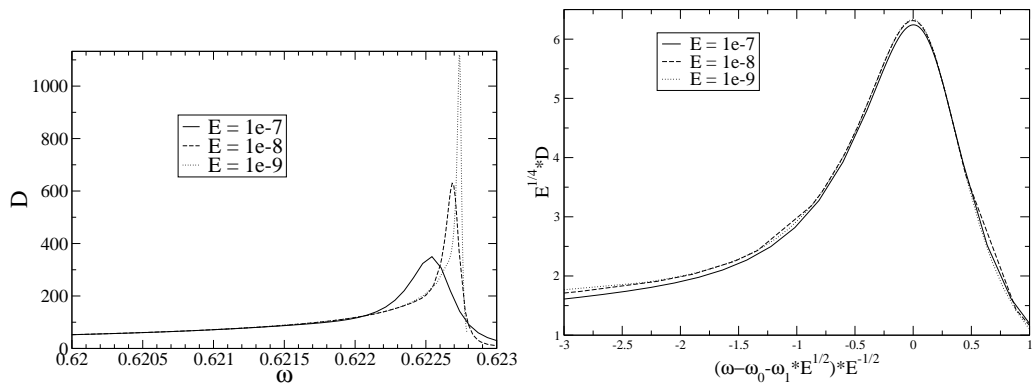

\centerline{ \includegraphics[width=0.5\linewidth,clip=true]{freq-diss_0.623.eps}
\includegraphics[width=0.5\linewidth,clip=true]{freq_normalize-diss_normalize_0.623.eps}
}
\caption[]{Left: Dissipation as a function of forcing frequency $\omega$
near the attractor of figure \ref{attractor}.  Note that far from the
resonance the dissipation tends to be independent of the Ekman number
$E$. Right: For the same attractor, scaled resonance curves. Note
that when scaled as indicated, dissipation curves no longer depend on
viscosity. Here $\omega_0=\omega_b$ is the upper bound of the frequency
interval of the attractor.}
\label{fig:freq-diss_0.623}
\end{figure}

To better understand the properties of the dissipation curves as shown
in figures~\ref{scan_tf} and \ref{scan_sym}, we shall first concentrate
on a given attractor, which is displayed in figure~\ref{attractor}. This
attractor may be found in the frequency range $[\omega_a,\omega_b]$
where  $\omega_b =\sqrt{\frac{5-\sqrt{5-4\eta}}{8}}$ ($\omega_a$ has no
analytic expression).
In figure~\ref{fig:freq-diss_0.623} we give a detailed view of the
resonance associated with this attractor around its asymptotic
frequency $\omega_b$.

\begin{figure}
         \centering
\includegraphics[width=0.6\linewidth,clip=true]{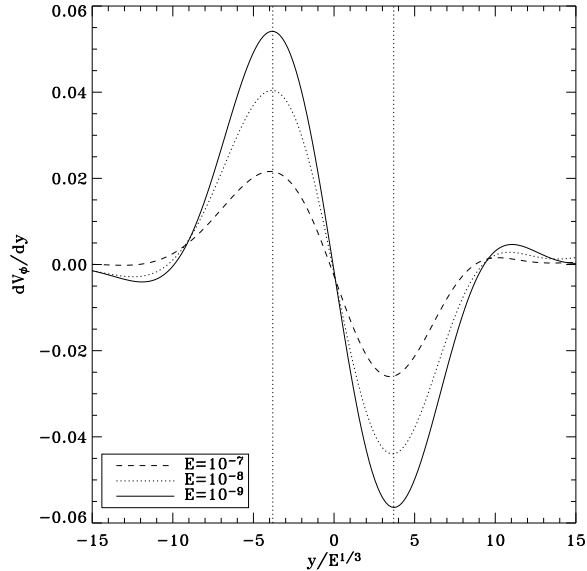}
\caption[]{
Profiles of the derivative $\frac{\dd v_\phi}{\dd y}$ along the straight
thick line shown in figure~\ref{attractor}-left crossing a segment of the
attractor at $\omega=0.621$, $\eta=0.35$.  The abscissa is normalised by
$E^{1/3}$: note that the location of the maxima of  $\frac{\dd v_\phi}{\dd
y}$ is nearly the same for the three profiles, showing that the width
of the layer scales with $E^{1/3}$.
}\label{profiles}
\end{figure}

The remarkable property of these curves is the independence of dissipation
from viscosity for frequencies far from the asymptotic frequency
$\omega_1$.  This result perfectly illustrates the demonstration of
\cite{O05} who showed that dissipation by inertial modes associated
with attractors with a finite Lyapunov exponent is actually independent
of the Ekman number. This comes from the width of the associated shear
layers,
which scale like $E^{1/3}$.  Figure~\ref{profiles} indeed confirms
that velocity gradients inside a shear layer vary on a scale changing
with the one-third power of the viscosity.

In \cite{RVG02} however, it was shown that shear layers associated with
freely decaying inertial modes, shaped by attractors, have a width
scaling like $E^{1/4}$. This means that at resonances, dissipation
resulting from the excitation of such singular modes diverges as
$E^{-1/4}$ in the asymptotic limit of small Ekman numbers (see below
sect.~\ref{simple_model}). Moreover, as
shown by \cite{RVG02}, the frequency of these eigenmodes is of the form
$\omega=\omega_0+\omega_1E^{1/2}$ for $E\ll1$.  We thus expect resonance
widths to be of the order of $E^{1/2}$.  Figure~\ref{fig:freq-diss_0.623}b
perfectly illustrates these properties, showing that if scaled properly,
resonance curves are independent of viscosity.

\subsubsection{Resonances associated with periodic
orbits}\label{section:periodic_orbits}

\begin{figure}
\centerline{ \includegraphics[width=0.6\linewidth,clip=true]{freq-diss_0.866_omegastretched.eps}
\includegraphics[width=0.4\linewidth,clip=true]{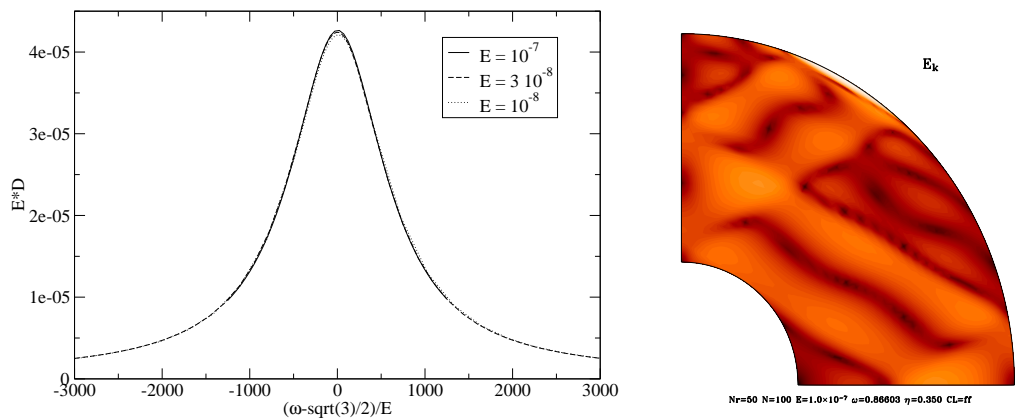} }
\caption[]{Left: Rescaled dissipation as a function of the
forcing frequency, also rescaled, near the periodic orbits with
$\theta_{cl}=\pi/3$ or $\omega=\sqrt{3}/2$. Right: the kinetic energy
distribution of the corresponding forced flow at $E=10^{-9}$.}
\label{fig:freq-diss_0.866_omegastretched}
\end{figure}

The foregoing resonances are not the strongest, however. Indeed,
we have shown in \cite{RGV01}, that there exist a finite number of
frequencies of the spectrum which are associated with strictly periodic
orbits of the characteristics. These frequencies are $\omega_{p,q} =
\sin(p\pi/2q)$ where $p$ and $q$ are integers; they are such that the
angle between characteristics and the rotation axis is a rational
fraction of $\pi$. Not all the rationals are allowed however,
because characteristics must propagate periodically inside or outside the ``shadow"
of the inner shell \cite[see][]{RGV01}. This conditions imposes that
$\eta\leq\omega\leq\sqrt{1-\eta^2}$ for the simplest periodic orbits
(it is more restrictive for more complex orbits). In the case chosen
here, namely $\eta=0.35$, the allowed periodic orbits are associated
with the three angles: $\pi/6$, $\pi/4$ and $\pi/3$. For the associated
frequencies the trajectories of characteristics are strictly periodic,
thus no shear layer is generated (the mapping has no focusing power).
This situation is illustrated in
figure~\ref{fig:freq-diss_0.866_omegastretched}. As shown, no small scale
comes in, and the dissipation diverges as $E^{-1}$ when $E\tv0$, while
the width of the resonances diminishes as $E$. Inspecting the spectral
content of the flow, we note that the critical latitude contributes to
some parts of the flow but at such a low level (less than $10^{-4}$)
that it does not influence the resonance.

\begin{figure}
\centerline{ \includegraphics[width=0.5\linewidth,clip=true]{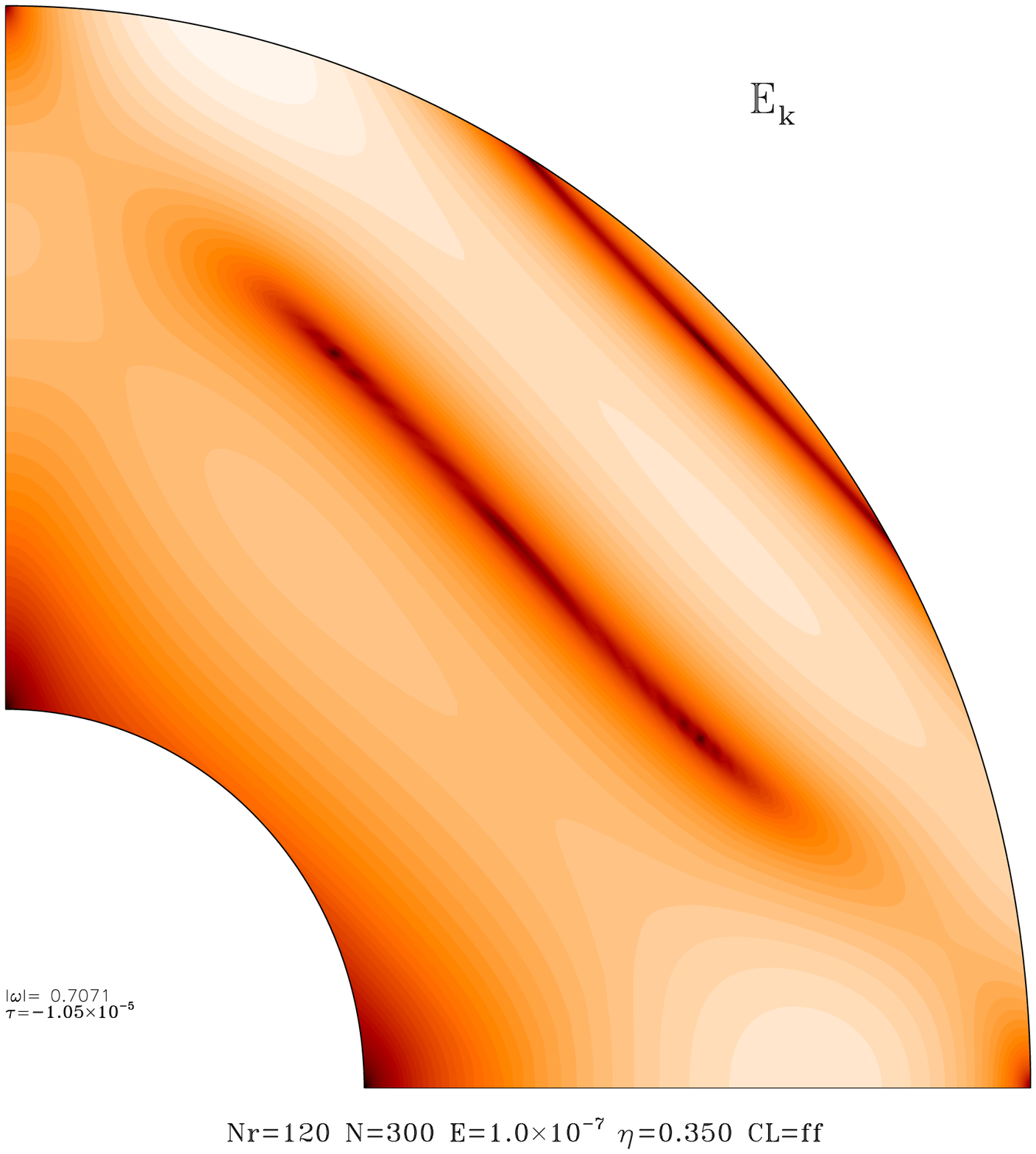}
\includegraphics[width=0.5\linewidth,clip=true]{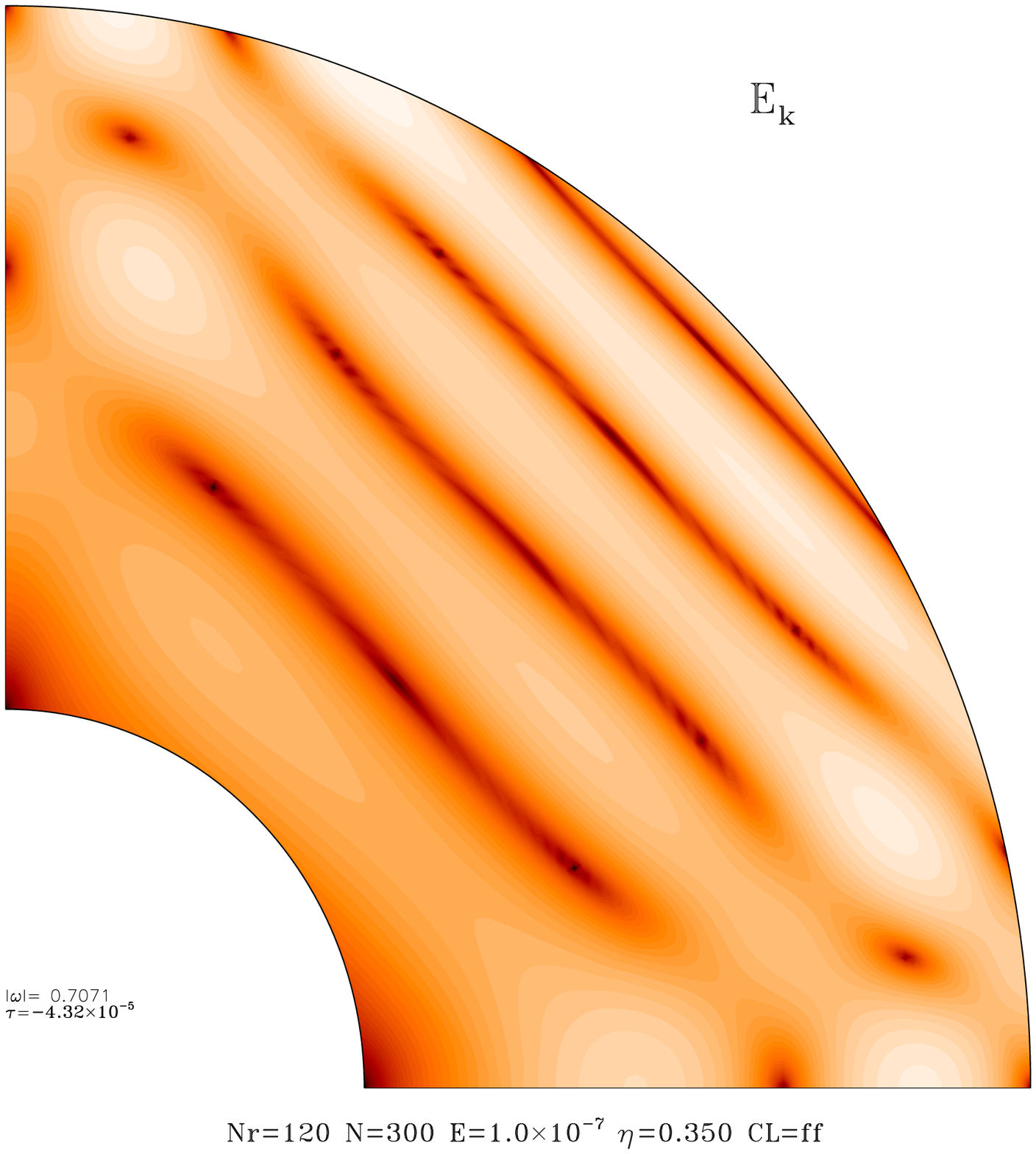} }
\caption[]{Kinetic energy distribution of the two first modes associated
with the $\vartheta=\pi/4$ periodic orbit given in table~\ref{tab_reg}.}
\label{mode_reg}
\end{figure}

This feature of the spectrum may be understood as follows. \cite{ML95}
showed that the two-dimensional semi-elliptic basin owns a denumerable
set of regular eigenmodes. These eigenmodes are associated with periodic
orbits of characteristics. They also showed that the eigenfrequencies
are infinitely degenerate. This is because of the ill-posed nature of
the eigenvalue problem: For each eigenvalue, the eigenmode is defined by
an arbitrary function which is given on the so-called fundamental
intervals of the boundary \cite[see][for details]{ML95}. In the case of
our fluid domain, the presence of an inner core removes almost all the
periodic orbits, letting only a finite number of them, depending on the
size of the core. However, for each of these frequencies which are of
the form $\sin(p\pi/q)$ (see above), the associated eigenfunctions are
still infinitely degenerate since they are specified by an arbitrary
function. We illustrate this property by computing the first eigenmodes
associated with these orbits (see table~\ref{tab_reg} and
figure~\ref{mode_reg}).

The response of the fluid to a time-periodic forcing near these
frequencies thus results essentially from the superposition of these
eigenmodes according to their projection on the exciting body force. As
the tidal forcing is on the large scale, the responding flow is
essentially on the large scales.

\begin{table}
\begin{center}
\begin{tabular}{|c|cc|c|cc|}
            & \multicolumn{2}{c|}{$\vartheta=\pi/6$} &
\multicolumn{1}{c|}{$\vartheta=\pi/4$} & \multicolumn{2}{c|}{$\vartheta=\pi/3$} \\
           & $\omega_2$ & $\tau_1$     &   $\tau_1$     & $\omega_2$ & $\tau_1$ \\
Mode 1     & $1.6\,10^8$& $-2.8\,10^3$ &   $-1.05\,10^2$& $-2\,10^{8}$ & $-2.8\,10^3$ \\
Mode 2     & $5.8\,10^9$& $-1.5\,10^4$ &   $-4.32\,10^2$& $-8\,10^{9}$ & $-1.5\,10^4$ \\
Mode 3     & $3.8\,10^{10}$& $-3.8\,10^4$ &  $-9.64\,10^2$& $-6\,10^{10}$ & $-3.7\,10^4$ \\
\end{tabular}
\caption[]{Eigenvalues of the first modes associated with
periodic orbits of the torus and verifying the same symmetry
as the forcing \eq{nat_force}. Each eigenvalue is written
$\lambda=i\sin\vartheta+i\omega_2E^2+\tau_1 E$, as expected for
regular modes with stress-free boundary conditions when $\omega_1=0$
(the reason for this vanishing term is not clear). The $\omega_2$-term
for the $\pi/4$-modes could not be evaluated because of round-off errors.
The coefficients $\tau_1$ and $\omega_2$ have been evaluated numerically
using values of $E$ around $10^{-7}$.}
\label{tab_reg}
\end{center}
\end{table}

\subsubsection{Dissipation at resonances}\label{simple_model}

The foregoing resonances, either associated with attractors (section
\ref{section:attractors}) or with periodic
orbits (section \ref{section:periodic_orbits}), may be described by a
simple model of a resonating eigenmode. Although much simplified, this
model is useful to understand the origin of the scaling laws verified by
the viscous dissipation.

Let $\calL$ be the linear operator governing the forced flow; we write

\beq i\omega\vu = \calL(\vu) + \vf \eeqn{eqf}
where $\omega$ is the frequency of the forcing. Let
$\la\vu_n\ra_{n\in\bbbn}$ be a set of eigenfunctions of $\calL$
verifying the same boundary conditions as $\vu$. Hence

\[ \lambda_n\vu_n = \calL(\vu_n)\]
where $\lambda_n=i\omega_n+\tau_n$ is the associated eigenvalue. We
assume that the $\la\vu_n\ra_{n\in\bbbn}$ form a complete orthogonal
basis. As far as modes associated with an attractor are concerned, we
have shown in \cite{RVG02}, that they may be described by a Hermite
function. They thus form a complete basis for the 1D functions defined
along a line orthogonal to the attractor. Thus we may write:

\[ \vu = \sum_na_n\vu_n, \qquad \vf = \sum_n f_n\vu_n\]
Here, we assumed that $\vf$ belongs to the same function space as
$\vu$. From \eq{eqf}, we easily find that $a_n=f_n/(i\omega-\lambda_n)$
so that

\beq \vu = \sum_{n} \frac{f_n\vu_n}{i\omega-\lambda_n} \eeqn{expun}
Now assuming that a single eigenmode dominates the series, we simplify
\eq{expun} as

\[ \vu \simeq \frac{f_n\vu_n}{i\omega-\lambda_n}\]
The dissipation rate can now be evaluated from (\ref{eq:dissip}).
The width of the shear layers is assumed to scale like $E^\alpha$.
The volume that contains the shear layers also scales like $E^{\alpha}$
and the gradients of the eigenmode $\vu_n$ scale like $E^{-\alpha}$.
We get:
\beq
D=\frac{E}{2}\intvol|c|^2 dV\sim
\frac{E^{1-\alpha}}{(\omega-\omega_n)^2+\tau_n^2}
\label{eq:Dtheorique}
\eeq
Assuming that the complex eigenfrequency of the excited mode expresses
as\linebreak\mbox{$\lambda_n=i\omega_n+\lambda_1E^\beta$}, we find that
\[
D\sim E^{1-\alpha-2\beta}
\]
at the resonance. For resonances associated with attractors with
vanishing Lyapunov exponents, the solutions of \cite{RVG02}
give $\alpha=1/4$, $\beta=1/2$ and thus $D\sim E^{-1/4}$ as observed
numerically. For the sharp resonances associated with the periodic
orbits of section \ref{section:periodic_orbits}, no small scales comes
in so that $\alpha=0$. The damping rate of a mode with a typical scale
independent of viscosity
is proportional to $E$; thus $\beta=1$ and we get $D\sim E^{-1}$
as also observed.

Although not fully rigorous, this short analysis shows that the observed
resonances of the two-dimensional model behave in a standard way.

\subsection{The critical latitude singularity}

Before ending this section we wish to discuss shortly the role played
by the critical latitude singularity. We recall that this singularity comes
from the ``oblique nature" of the boundary conditions to be used with the
Poincaré equation \cite[see][]{RGV01}. It leads to a singularity of the
solutions but weaker than the one associated with the attractors. When
the problem takes into account the fluid's viscosity, this singularity
manifests itself as a broadening of the Ekman layer, which thickens
from the usual $E^{1/2}$ scale to the $E^{2/5}$ scale, on a latitudinal
extension that is \od{E^{1/5}} \cite[][]{RS63}. As illustrated in
figure~\ref{attractor} this singularity generates its own network of
shear layers thus adding some dissipation to the one of the attractor.
However, the contribution of this singularity is much smaller than
the one of the attractor.  We also observe that if the frequency of
the forcing is not such that the attractor has a branch grazing at the
critical latitude, the amplitude of the flow in this region vanishes
with a vanishing Ekman number.

\subsection{Summarising the two-dimensional case}

The preceding results show that the dissipation associated with
periodically forced shear layers may vary very strongly as a function of
the frequency of the forcing. At a generic frequency, we find that the
dissipation is independent of the viscosity and thus confirm the analysis
of \cite{O05}. However, the inertial frequency band contains also
infinitely many frequencies (at accumulation points) where attractors
are weaker (their Lyapunov exponent vanishes and the convergence
of characteristics is algebraic instead of exponential). At these
points the dissipation strongly depends on viscosity, namely as
$E^{-1/4}$. Finally, we also exhibited resonances that are associated
with the few allowed periodic orbits of characteristics. These resonances
are those of the regular modes which remain of the dense set of modes of
the $\eta=0$ torus \cite[see][]{ML95}.
As these periodic orbits, there is only a finite
number of such resonances for a given set-up with $\eta\neq0$.

We now examine the three-dimensional case so as to determine which of
these properties remain in this more realistic case.

\section{Resonances in the spherical shell}

\subsection{Numerics}

Turning to the spherical shell problem, we now solve \eq{eqmotion} in
spherical geometry. We discretize the unknowns and the equations using an
expansion of the fields on the spherical harmonics $\YL(\theta,\varphi)$
for the horizontal part and using the Chebyshev polynomials on the
Gauss-Lobatto collocations nodes for the radial part.  Details may be
found in \cite{RV97}. We just recall here that the velocity field is
expanded as

\[\vu=\sum_{l=0}^{+\infty}\sum_{m=-l}^{+l}\ulm(r)\RL+\vlm(r)\SL+\wlm(r)\TL
,\]
with

\[\RL=\YL(\theta,\varphi)\vec{e}_{r},\qquad \SL=\na\YL,\qquad
\TL=\na\times\RL \]
where gradients are taken on the unit sphere. Using the same expansion
for the body force \eq{force}, we find

\[ \vf = \calA r^2\lp -i\omega\vR_2^0 + \frac{1}{\sqrt{15}}\vT_1^0 -
\frac{12}{\sqrt{35}}\vT_3^0\rp\]
Following \cite{rieu87}, we derive the equations for the radial
functions $\ulm(r)$ and $\wlm(r)$ from the equations of vorticity and
continuity. They read

\greq
E\Deltal\wl - i\omega \wl= \\
 \hspace{1.5cm} -A_{\ell}r^{\ell-1}\dr{} \biggl( \frac{\ulmm}{r^{\ell-2}}\biggr)
-A_{\ell+1}r^{-\ell-2}\dr{}\biggl( r^{\ell+3}\ulp\biggr)
- r^2(\frac{\delta_{\ell,1}}{\sqrt{3}}-\frac{12}{\sqrt{7}}\delta_{\ell,3}) \\
\\
E\Deltal\Deltal(r\ul)-i\omega \Deltal(r\ul)= \\
\hspace{1.5cm} B_{\ell}r^{\ell-1}\frac{\partial}{\partial r}
\biggl(\frac{\wlmm}{r^{\ell-1}}\biggr) + B_{\ell+1}r^{-\ell-2}
\dr{}\biggl( r^{\ell+2}w^{\ell+1}\biggr)- i\omega6\sqrt{5} r\delta_{\ell,2}
\egreqn{eqproj}
where axisymmetry has been assumed. We have used

\[ A_\ell = \frac{1}{\ell\sqrt{4\ell^2-1}}, \qquad
B_\ell = \ell^2(\ell^2-1)A_\ell, \qquad \Deltal = \frac{1}{r}\ddnr{}r -
\frac{\ell(\ell+1)}{r^2} \]
The set of equations \eq{eqproj} is completed by stress-free boundary
conditions, which read:

\[ \ulm = \ddr{r\ulm} = \dr{}\lp\frac{\wlm}{r}\rp =0 \]
for the radial functions taken at $r=\eta$ or $r=1$.

\begin{figure}
\centering
\includegraphics[width=\linewidth,clip=true]{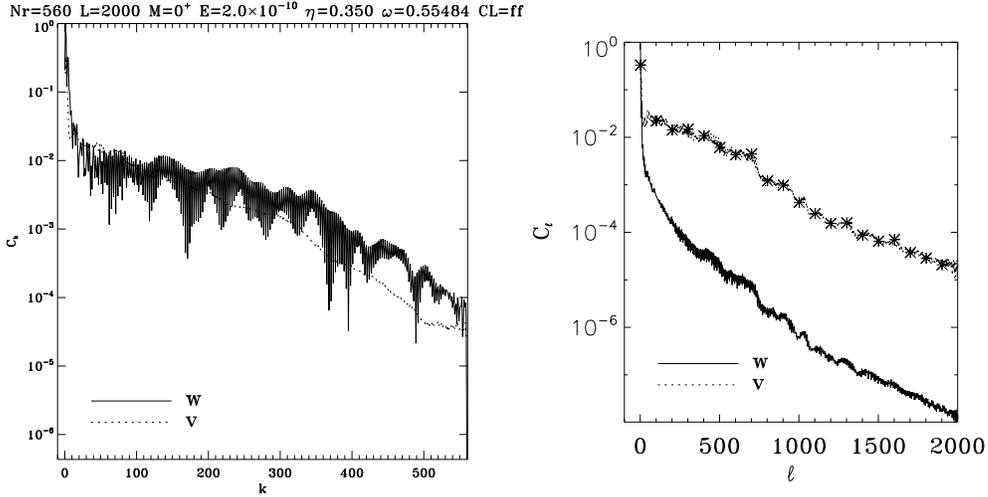}
\caption[]{Spectral content of the solution shown in
figure~\ref{zoom_diss555} middle bottom.  Left: Maximum Chebyshev spectrum
of the velocity field: for a given order of the Chebyshev coefficient
we plot the maximum absolute value of the coefficient obtained over all
the spherical harmonic components. Right: The maximum spherical harmonic
spectrum of the velocity field: for each spherical harmonic order $\ell$
we plot the largest absolute value of all the Chebyshev coefficients;
$w$ refers to $\wl$ while $v$ refers to $r\ul$ \cite[see also][]{RV97}.
}\label{spectra}
\end{figure}

In figure~\ref{spectra}, we give an example of the spectral content of
the forced flow shown in figure~\ref{zoom_diss555} at bottom centre,
for $E=2\,10^{-10}$. The spectra show that the truncation error of the
solutions are \od{10^{-4}}, while we estimated the round-off error (with double precision arithmetics)
to a lower value, in this case \cite[see][for a more thorough discussion of these error
matters]{VRBF07}.

\subsection{Overall view of the resonance spectrum}

\begin{figure}
\centering
\includegraphics[width=\linewidth,clip=true]{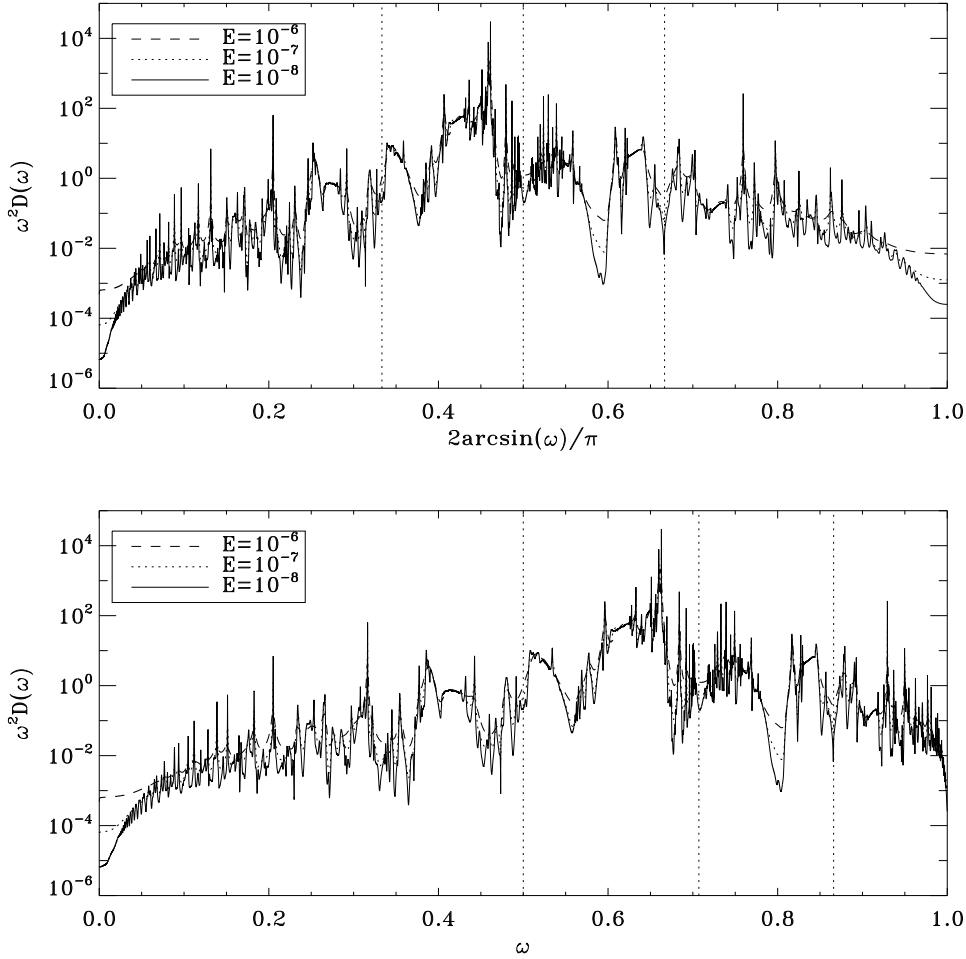}
\caption[]{Scan of the whole inertial band for the viscous dissipation in
a spherical shell.  Vertical dotted lines mark the frequencies allowing
strictly periodic orbits of characteristics $\vartheta=\pi/6, \pi/4,
\pi/3$.}
\label{scan_diss}
\end{figure}

As for the two dimensional case, we first scan the whole inertial
band, computing the viscous dissipation. The result is plotted in
figure~\ref{scan_diss}. There we note that the response curve is
very spiky as in the two-dimensional case, revealing very strong
variations (six orders of magnitude at E=$10^{-8}$). We plotted the
quantity $\omega^2D(\omega)$ to remove the $1/\omega^2$-divergence
at low frequencies, since in this range, the flow is dominated by the
velocity field $\vu = Ar^2\sth/\omega \ephi$.  The broad shape of the
curve does not show a marked symmetry with respect to $\vartheta=\pi/4$
unlike its two-dimensional counter-part (figure~\ref{scan_tf}). This
is expected since the curvature terms of the 3D-operators contribute
to break this symmetry. Another striking difference is the absence of
resonances at the three frequencies corresponding to strictly periodic
orbits of characteristics, namely $\vartheta=\pi/6, \pi/4, \pi/3$. On
the contrary, there is an anti-resonance phenomena.

\begin{figure}
\centering
\includegraphics[width=\linewidth,clip=true]{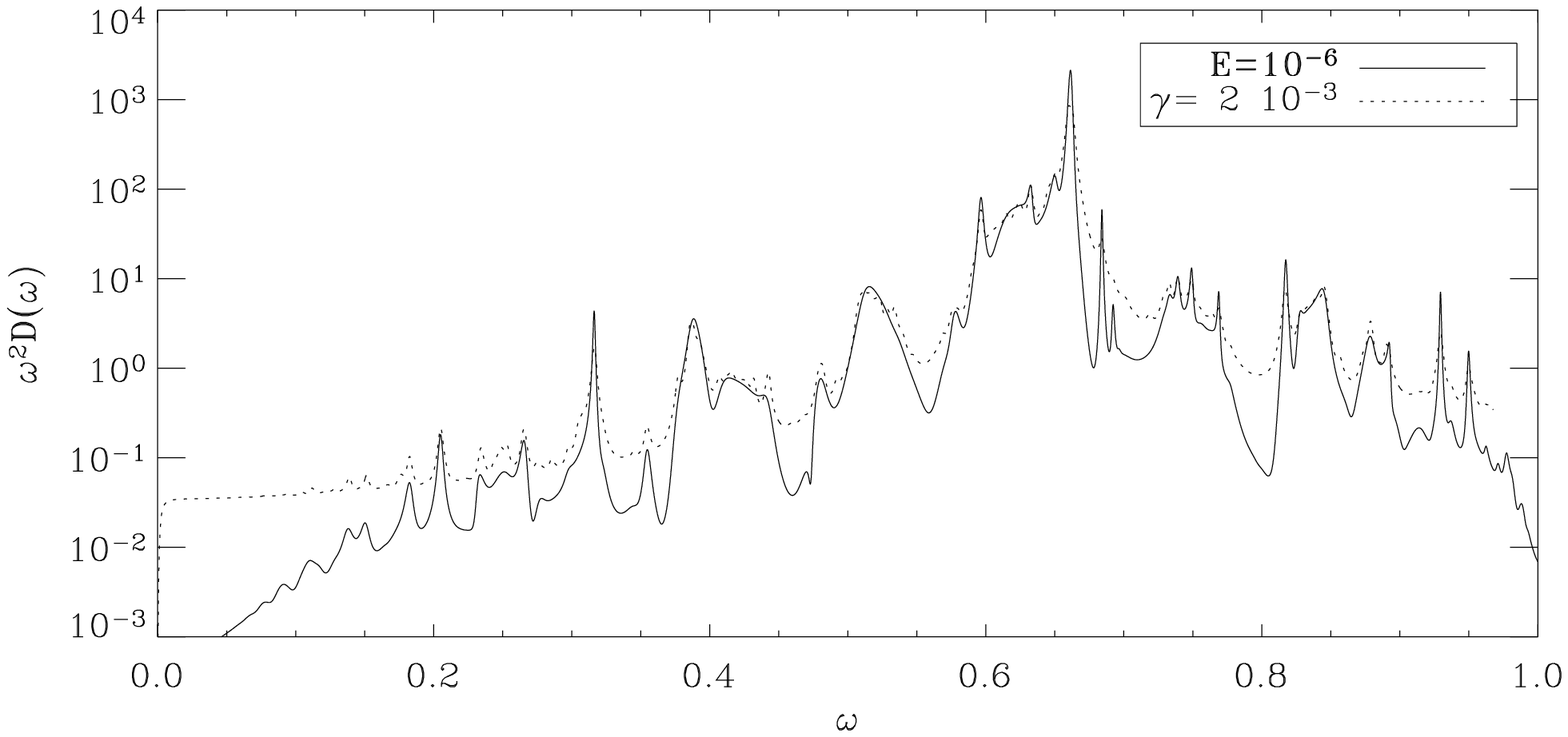}
\caption[]{Scan of the whole inertial band for the dissipation using the
viscous force (solid line) or the friction force (dotted line). Note the similarity of the curves.}
\label{scan_fric}
\end{figure}

Before discussing in more details some specific features of this curve,
let us focus on figure~\ref{scan_fric}. Here, following the idea of
\cite{O09}, we have plotted the dissipation curve when a frictional
force $-\gamma\vu$ replaces the viscous force $E\Delta\vu$.  As shown by
figure~\ref{scan_fric}, the two curves are very similar, demonstrating
that the fluid response is mainly governed by the underlying Poincaré
operator (i.e. the operator governing the inviscid problem). The meaning
of this similarity and the relation with the Poincaré operator may be
enlighted when using this simple frictional force. In this case the
flow verifies:

\greq
(i\omega+\gamma)\vu + \ez\times\vu  = -\na p +\vf\\
\\
\Div\vu = 0
\egreqn{eqfric}
Setting $\lambda=i\omega+\gamma$, this problem, which is completed by
the boundary condition $\vu\cdot\vn=0$, may be symbolically written:

\[ \lambda\lp\begin{array}{cc}{\rm Id} & 0 \\ 0&0 \\ 0&0\end{array}\rp \lp \begin{array}{c}\vu\\
p\end{array}\rp = \lp\begin{array}{cc} -\ez\times & -\na \\ \na\cdot & 0\\ \vn\cdot & 0\end{array}\rp
\lp \begin{array}{c}\vu\\ p\end{array}\rp + \lp \begin{array}{c}\vf \\0 \\0 \end{array}\rp\]
or

\[ \lp \lambda J - {\cal L}_P\rp \vX = \vX_f, \qquad \vX=\lp \begin{array}{c}\vu\\
p\end{array}\rp, \quad \vX_f = \lp \begin{array}{c}\vf \\0 \\0 \end{array}\rp, \quad
J=\lp\begin{array}{cc}{\rm Id} & 0 \\ 0&0 \\ 0&0\end{array}\rp\]
Here, $\vX=\;^t\!(\vu,p)$ and ${\cal L}_P$ symbolizes the Poincaré operator. 
Formally, the solution of the velocity field may be expressed as:

\[ \vu = (\lambda J - {\cal L}_P)^{-1}\vf\]
where $(\lambda J - {\cal L}_P)^{-1}$ is a kind of resolvent of the
Poincaré operator, restricted to a vector field, since the operator $J$
is ``close" to the identity.  The dissipation associated with the flow
is proportional to the norm of $\vu$; indeed, it is

\[ D=\gamma\intvol |\vu|^2dV\]
Hence, using a restriction of the norm of $\vX$ to the velocity field, we may
write the dissipation as

\[ D=||(\lambda J - {\cal L}_P)^{-1}\vf||^2\]
Now, let us consider the $\epsilon$-pseudospectrum of the Poincaré
operator. This quantity is indeed very appropriate to deal with a non-normal
operator such as the Poincaré one \cite[see][]{TE05}. It is defined
as the set of complex numbers $\lambda$ such that

\[ ||(\lambda J - {\cal L}_P)^{-1}|| > 1/\epsilon\]
where we recall that the norm of a bounded linear operator $L$ is the number

\[ \max_{\rm over\; \vX} \lp \frac{||L\vX||}{||\vX||}\rp\]
We therefore see that all the regions of the frequency axis where
$D>\gamma/\epsilon^2$ belong to the $\epsilon$-pseudospectrum of the
Poincaré operator. Indeed, if $D>\gamma/\epsilon^2$ then

\[ 1/\epsilon < ||(\lambda J - {\cal L}_P)^{-1}\vf|| \leq ||(\lambda J - {\cal L}_P)^{-1}||\]
where we assumed that $||\vf||=1$.

Thus, the dissipation curve obtained with the frictional force gives a
partial (one-dimensional) view of the $\epsilon$-pseudospectrum of the
Poincaré operator; computing the curve for many $\gamma$'s would give a
view of the two-dimensional subsets of the complex plane, which belong
to this pseudospectrum.

Now, the similarity of the two dissipation curves associated with the two
damping forces, comes from the nature of the solutions. When these are
in the form of a well defined attractor, the width of the shear layers
is a small scale that singles out so that $E\Delta\vu\sim-Ek^2\vu$,
where $k\sim E^{-1/3}$.

To conclude this point, we see that the solution of the forced
problem show another side of the Poincaré operator, namely its
pseudospectrum, and this quantity is independent of both the
forcing and the frictional force. The peaks of the dissipation curves
offer a partial (1D-) view of this quantity.

\subsection{Anti-resonances at periodic orbits of characteristics}
\label{antir}

We designate by anti-resonances of the dissipation curve, the frequencies
for which the dissipation vanishes with the Ekman number.

\begin{figure}
\centerline{
\includegraphics[width=0.5\linewidth,clip=true]{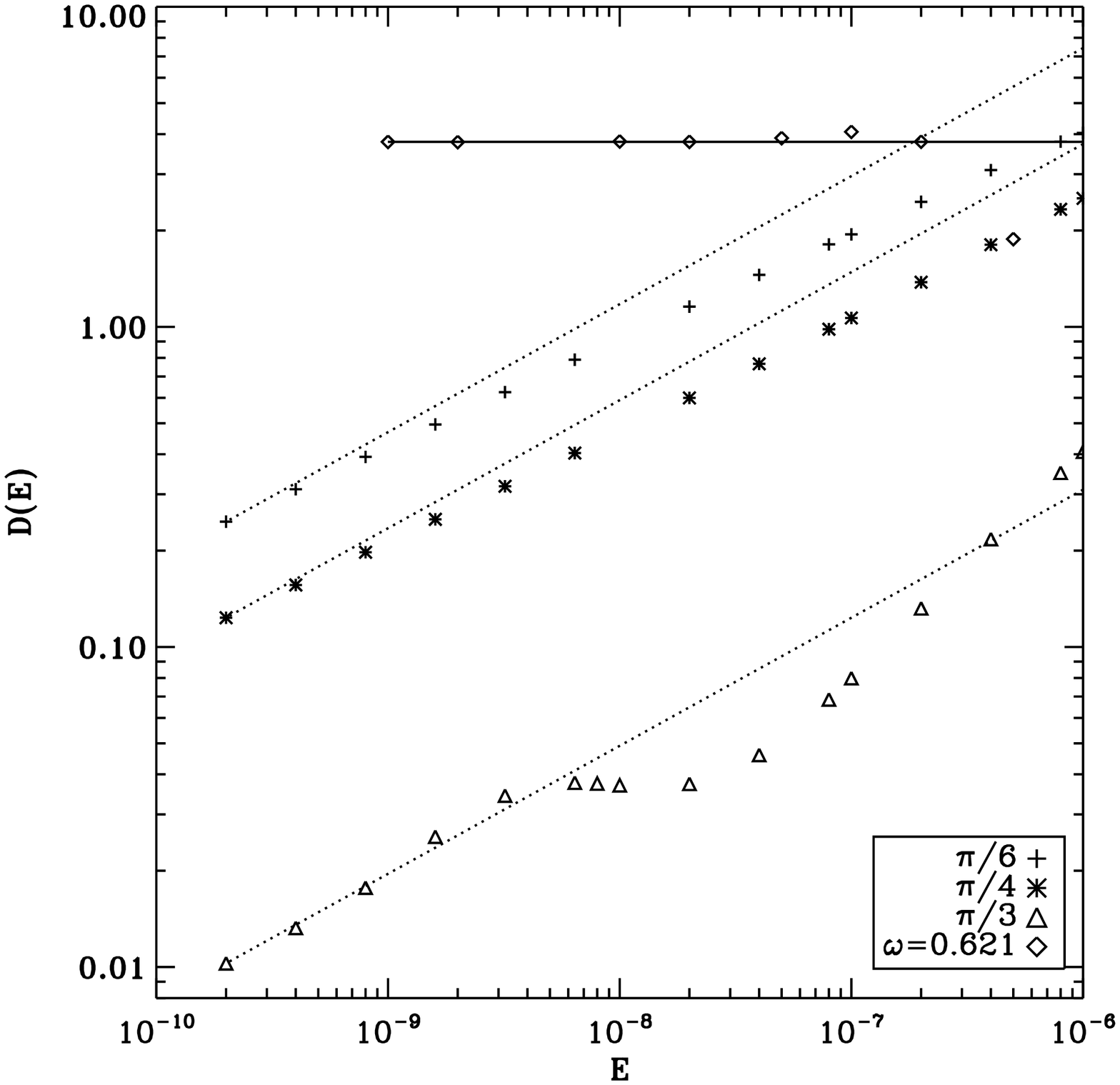}
\includegraphics[width=0.5\linewidth,clip=true]{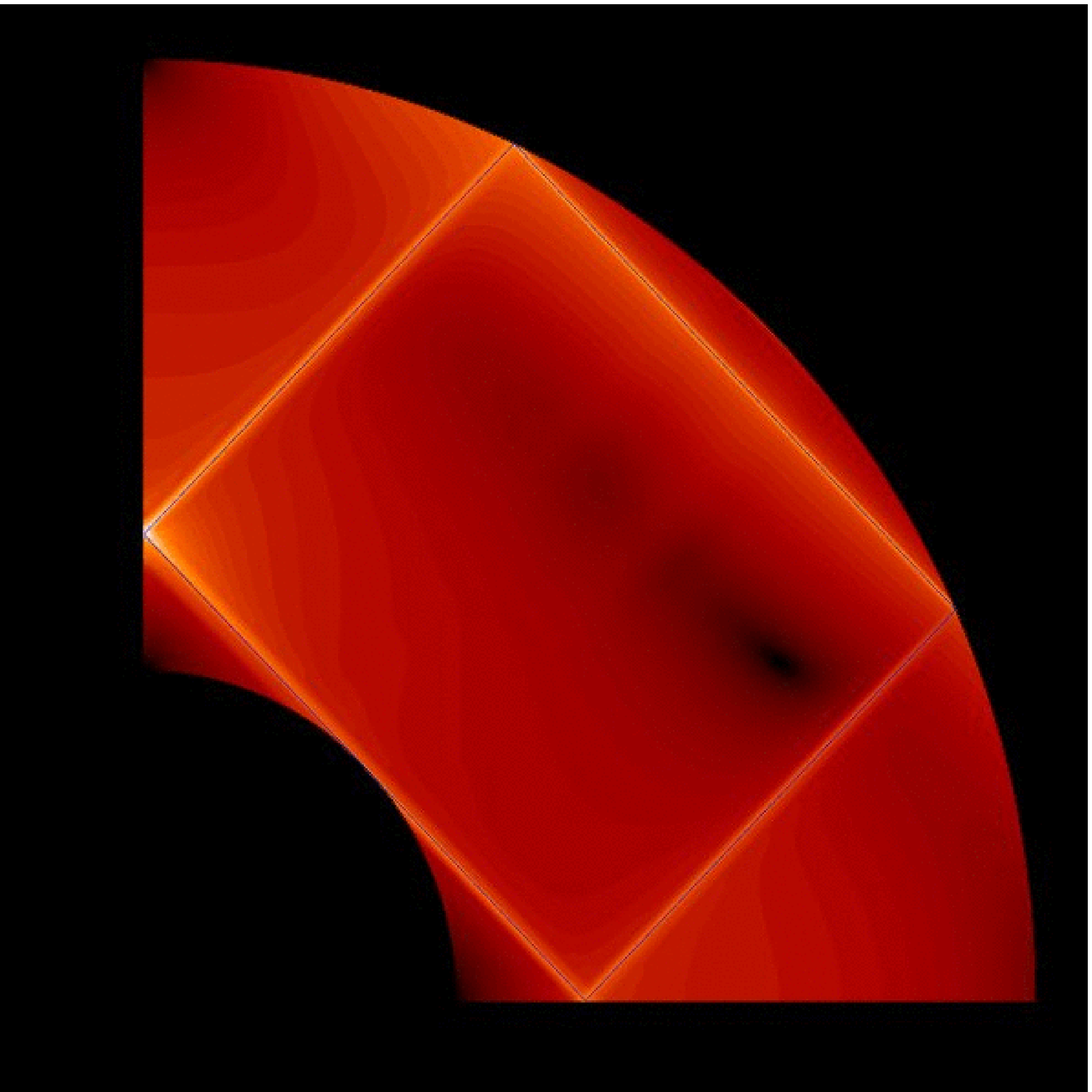}
}
\caption[]{Left: Dissipation as a function of the Ekman number for four
given frequencies.  'Diamonds': $|D(E)-D(E=10^{-6})|$ for a forcing
frequency exciting the attractor shown in figure~\ref{attractor} and
\ref{zoom_diss} at $\omega=0.621$. 'Pluses', 'Stars' and 'Triangles' show
the dissipation at frequencies respectively: $\sin(\pi/6)$, $\sin(\pi/4)$
and $\sin(\pi/3)$; the dotted straight lines emphasize the power law
$D(E)\propto E^{2/5}$. Right: the kinetic energy distribution for a
forcing at $\omega=\sin(\pi/4)$ and E=$2\,10^{-9}$.  }

\label{scl}
\end{figure}

We consider the three strictly periodic orbits authorised in the
shell with $\eta=0.35$, namely those orbits for which characteristics
remain either inside or outside the shadow of the inner core
\cite[see][]{RGV01}. These are associated with $\vartheta=\pi/6,
\pi/4, \pi/3$. We plot in figure~\ref{scl} the dissipation for these
three values and note that it decreases following quite closely the
law $D\propto E^{2/5}$. Such a scaling is obviously reminiscent of the
critical latitude boundary layer. Actually, as shown by figure~\ref{scl}
(right), the fluid's oscillation is confined along the characteristics
emitted by the critical latitude boundary layer.

One may retrieve the scaling of dissipation if we note that the volume
of the boundary layer surrounding the critical latitude singularity is
\od{E^{1/5+2/5}} and that the velocity field scales as $E^{-1/5}$. This
scaling appears if we observe that the singular inviscid field has
a finite normal velocity at the boundary and an infinite tangential
velocity \cite[see][]{RGV01}. Assuming that $u_r$ is of order unity
(as the forcing), boundary layers relations imply that the tangential
velocity scales as $E^{-1/5}$. Using the second expression of the dissipation,
$D=\intvol Re(\vu^*\cdot\vf) dV$, we recover the scaling law that
we observe numerically. We note that this regularised singularity,
propagates along the (periodic path of) characteristics, but without any
focusing, the resulting shear layer widens slowly as we move away from the
critical latitude of the inner bounding sphere.

\subsection{Dissipation in the frequency bands with strong attractors}

\begin{figure}
\centerline{
\includegraphics[width=0.5\linewidth,clip=true]{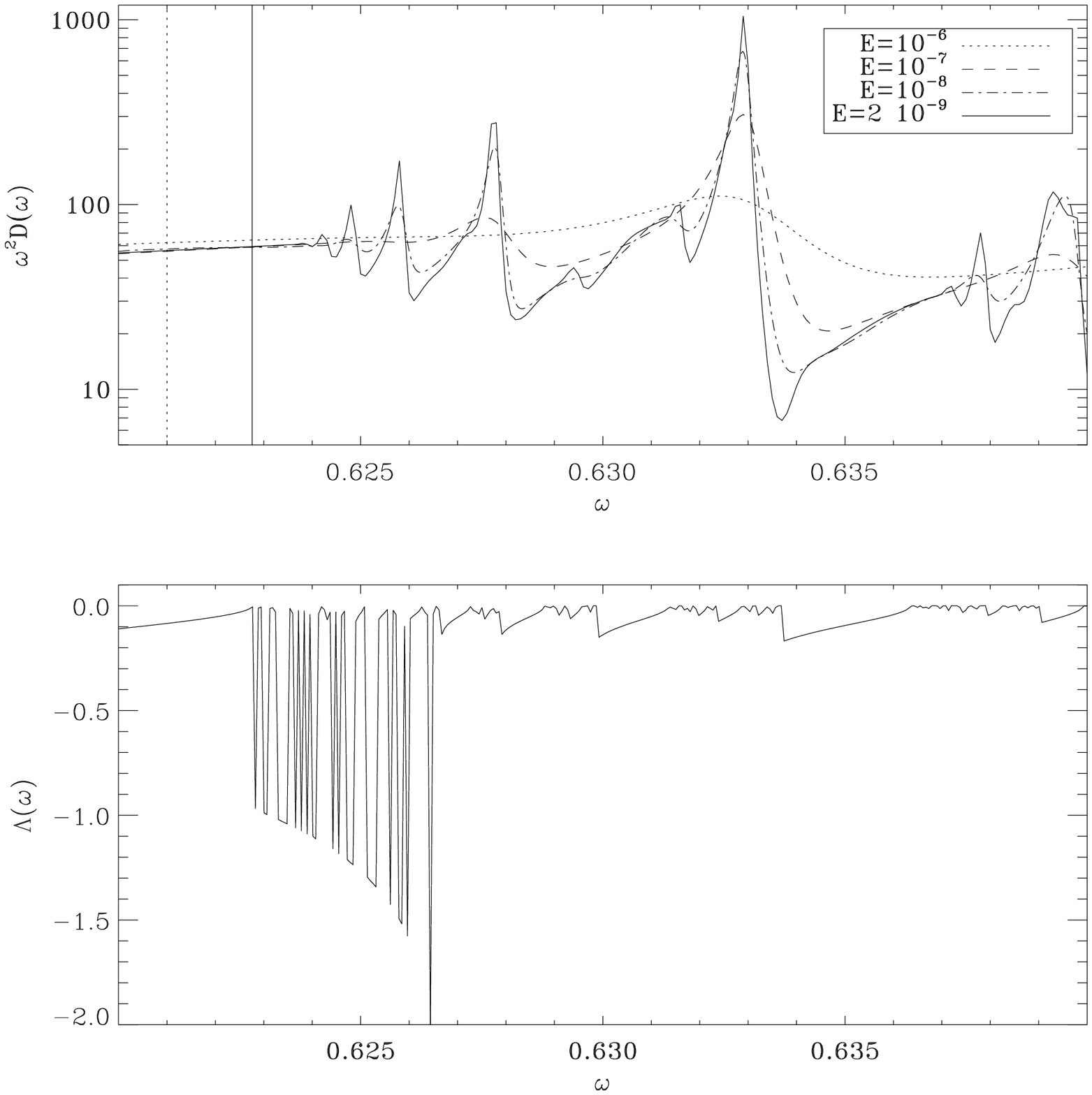}
\includegraphics[width=0.5\linewidth,clip=true]{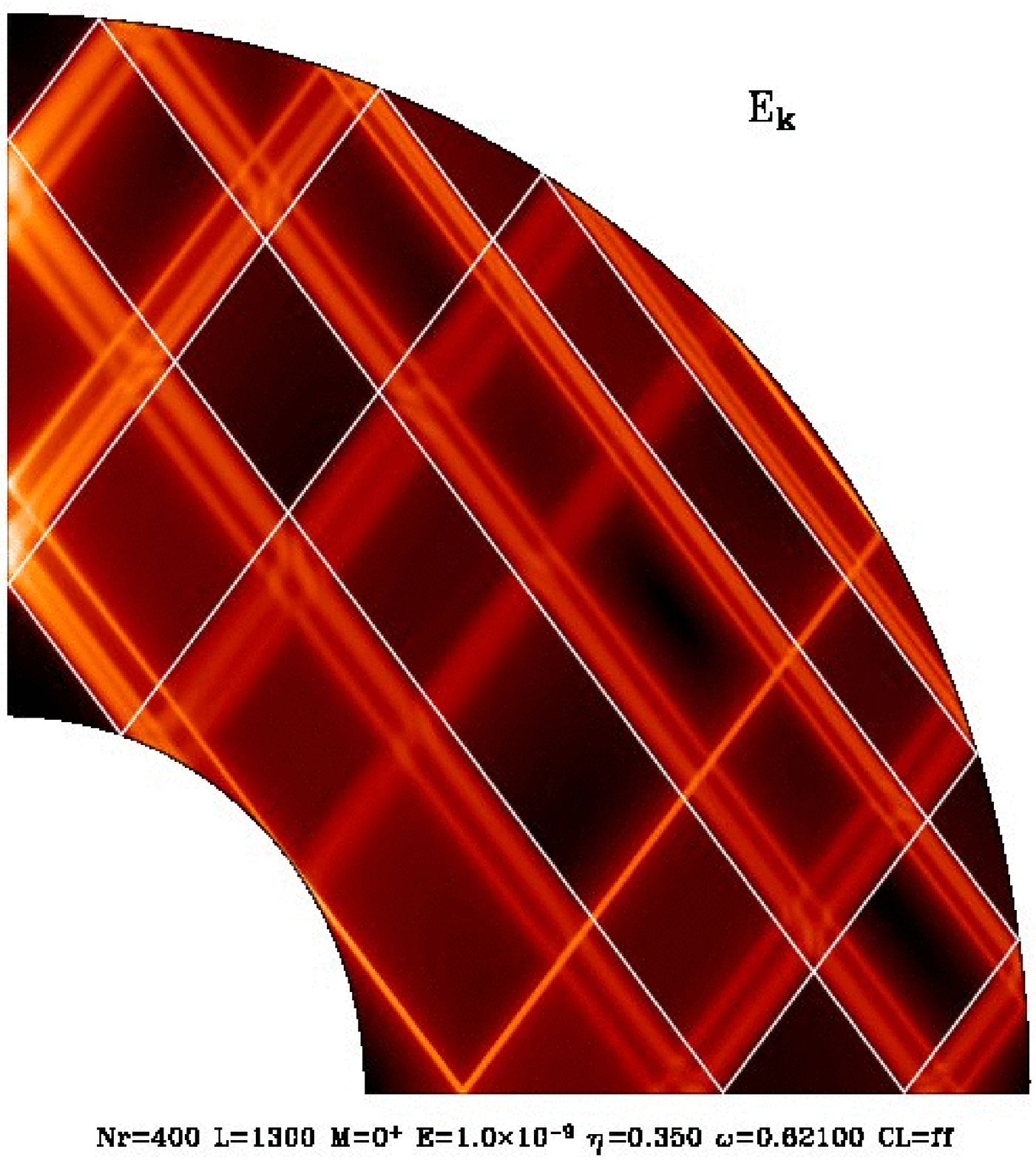}
}
\caption[]{Left (top): Zoom of the dissipation in the spherical shell
for $\omega\in[0.62, 0.64]$.  Left (bottom) : The associated Lyapunov
exponents in the same frequency range. Right: The kinetic energy of the
forced flow at $\omega=0.621$ viewed in a meridional plane. The white line marks the path of the
attractor.}
\label{zoom_diss}
\end{figure}

One of the main results established in two dimensions is that a forced
flow oscillating at the frequency of an attractor dissipates energy
independently of viscosity, provided it is low enough. As shown in
figure~\ref{zoom_diss}, which gives a zoom on the dissipation curve,
we clearly see a region of the spectrum (below $\omega=0.624$) where
the dissipation is independent of the Ekman number. The curve in
figure~\ref{scl} (diamonds) confirms this convergence to a finite
dissipation as viscosity vanishes.

There is however a surprise in this case. If we look at the Lyapunov
exponent in the same frequency band we note that something should happen at the
frequency $\omega_0$ where this exponent vanishes. According to the
results of \cite{RGV01}, we should expect a resonance, as in
two-dimensions. No such resonance occurs, and this is independent of the
symmetry of the forcing. Inspection of the associated flow (see
figure~\ref{zoom_diss} right) shows that the attractor is not excited. Rather, the
critical latitude singularity excites a shear layer, which propagates
towards the attractor. This is all the more surprising that in this
frequency band there are many eigenmodes well centered on the attractor
path (an example may be found in \citealt{RGV01}).

\begin{figure}
\centerline{\includegraphics[width=1.\linewidth,clip=true]{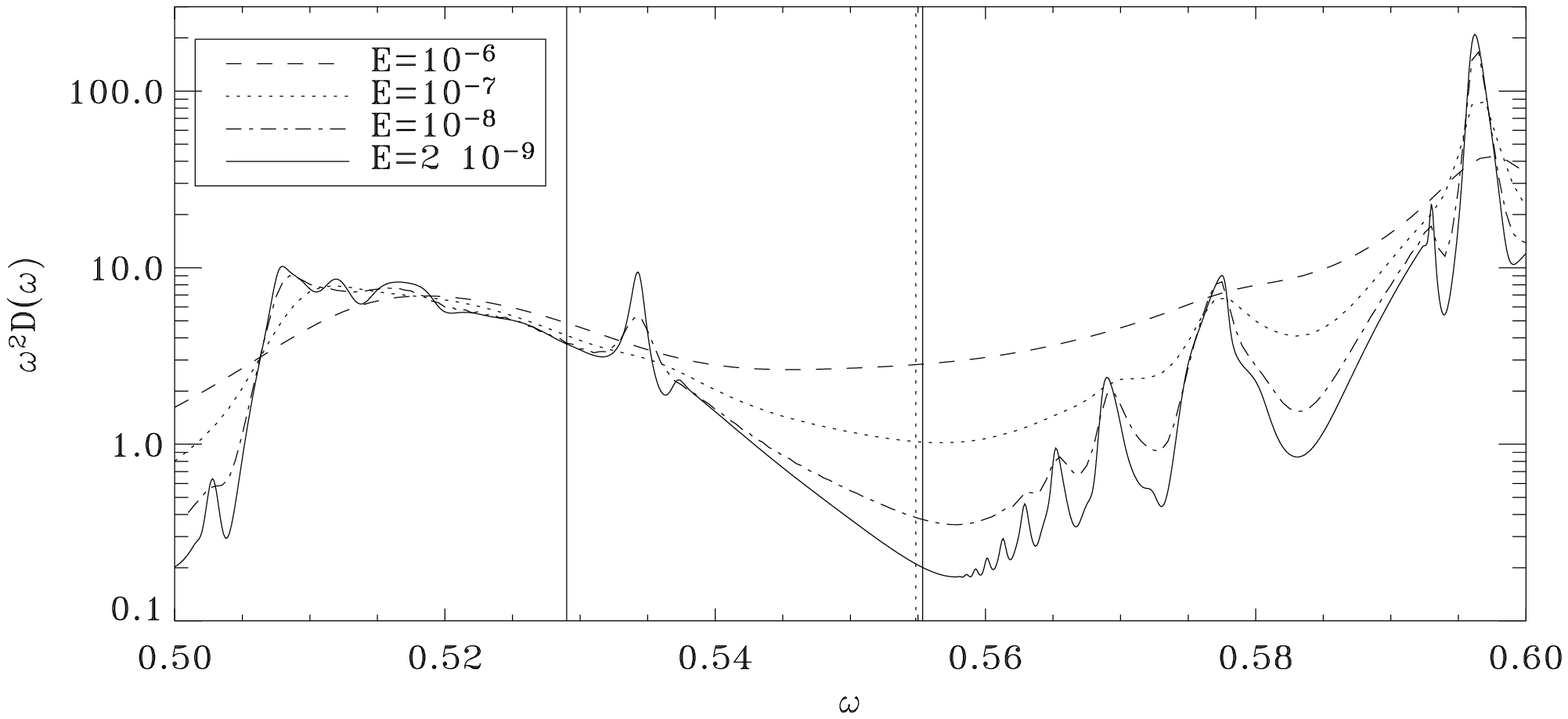}}
\centerline{
\includegraphics[width=0.33\linewidth,clip=true]{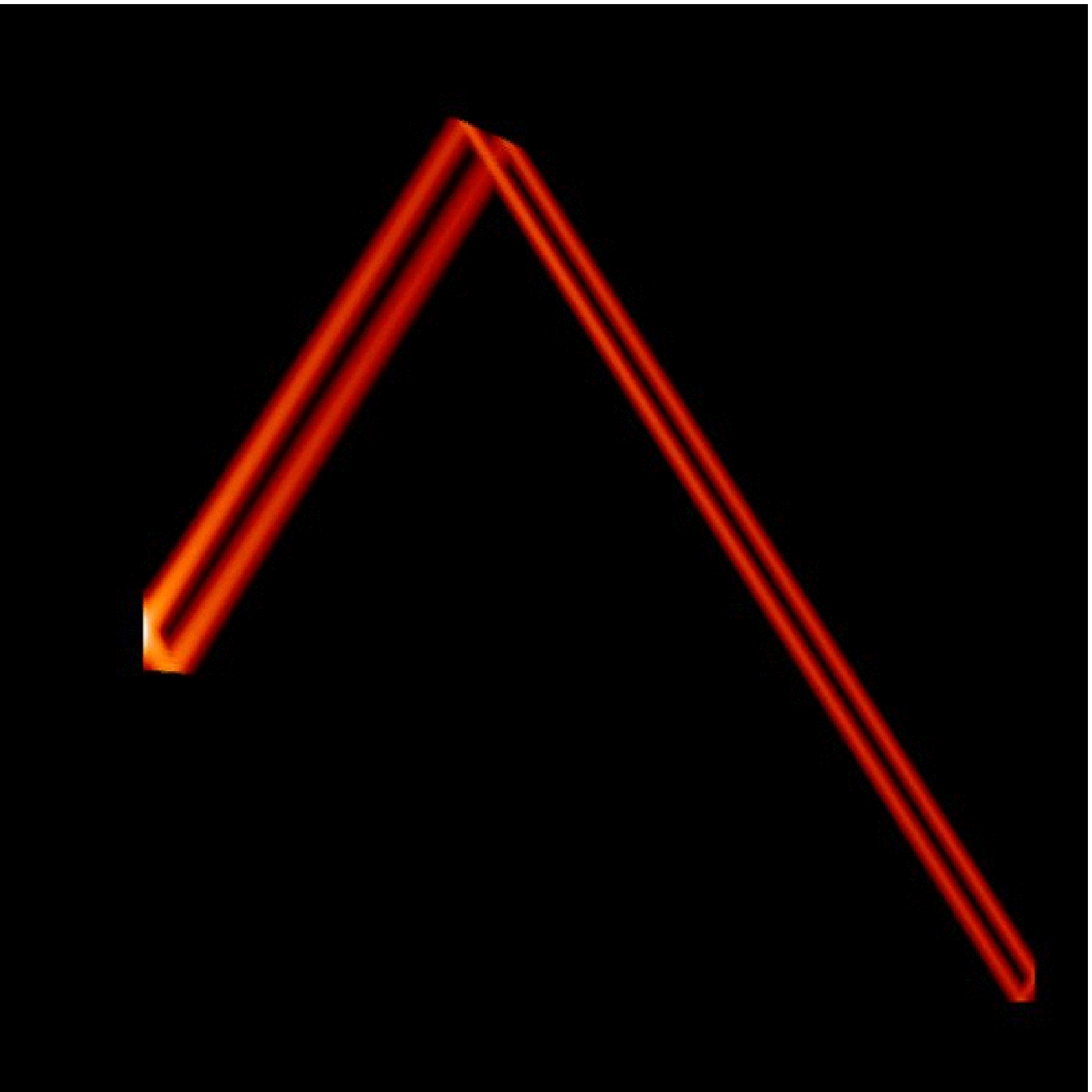}
\includegraphics[width=0.33\linewidth,clip=true]{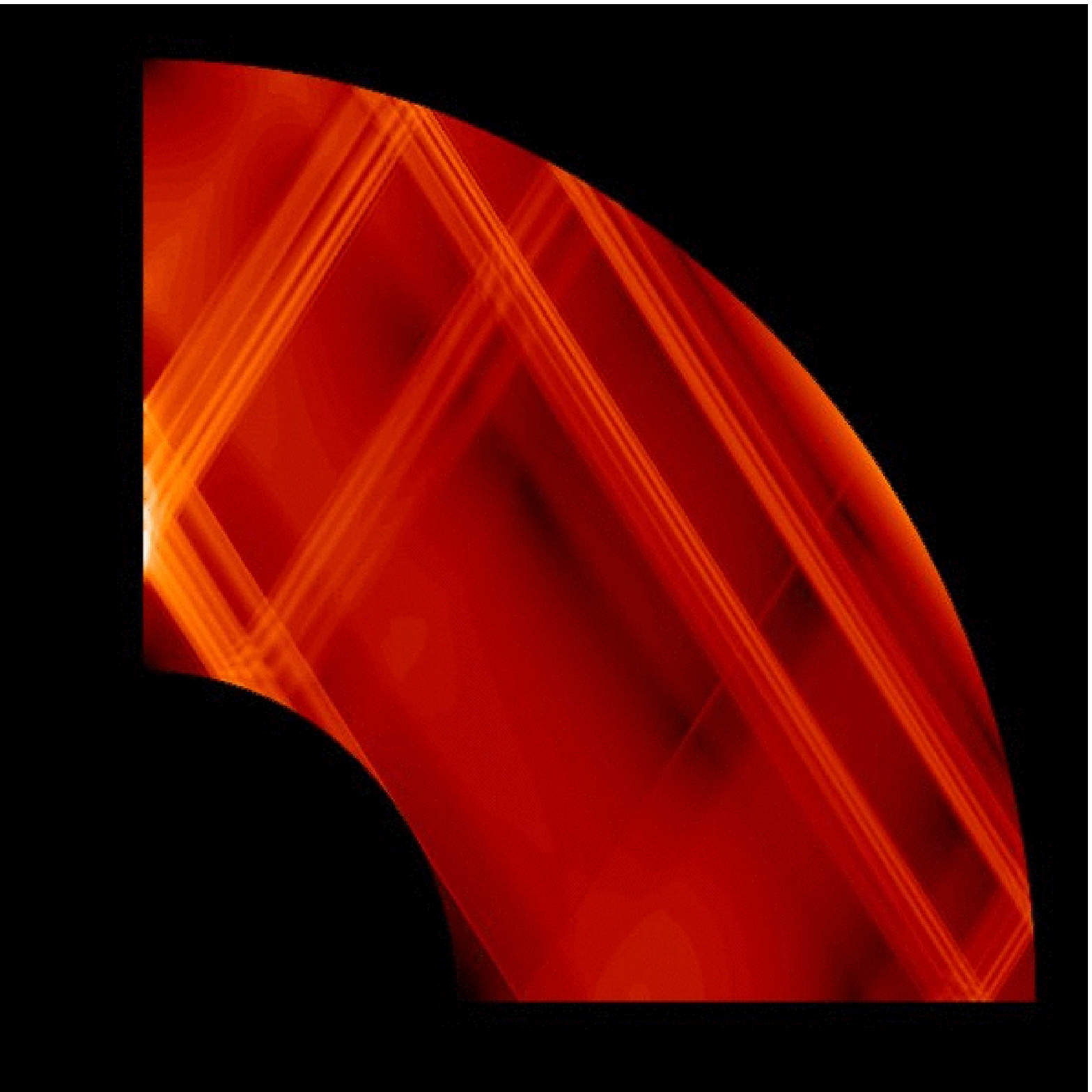}
\includegraphics[width=0.33\linewidth,clip=true]{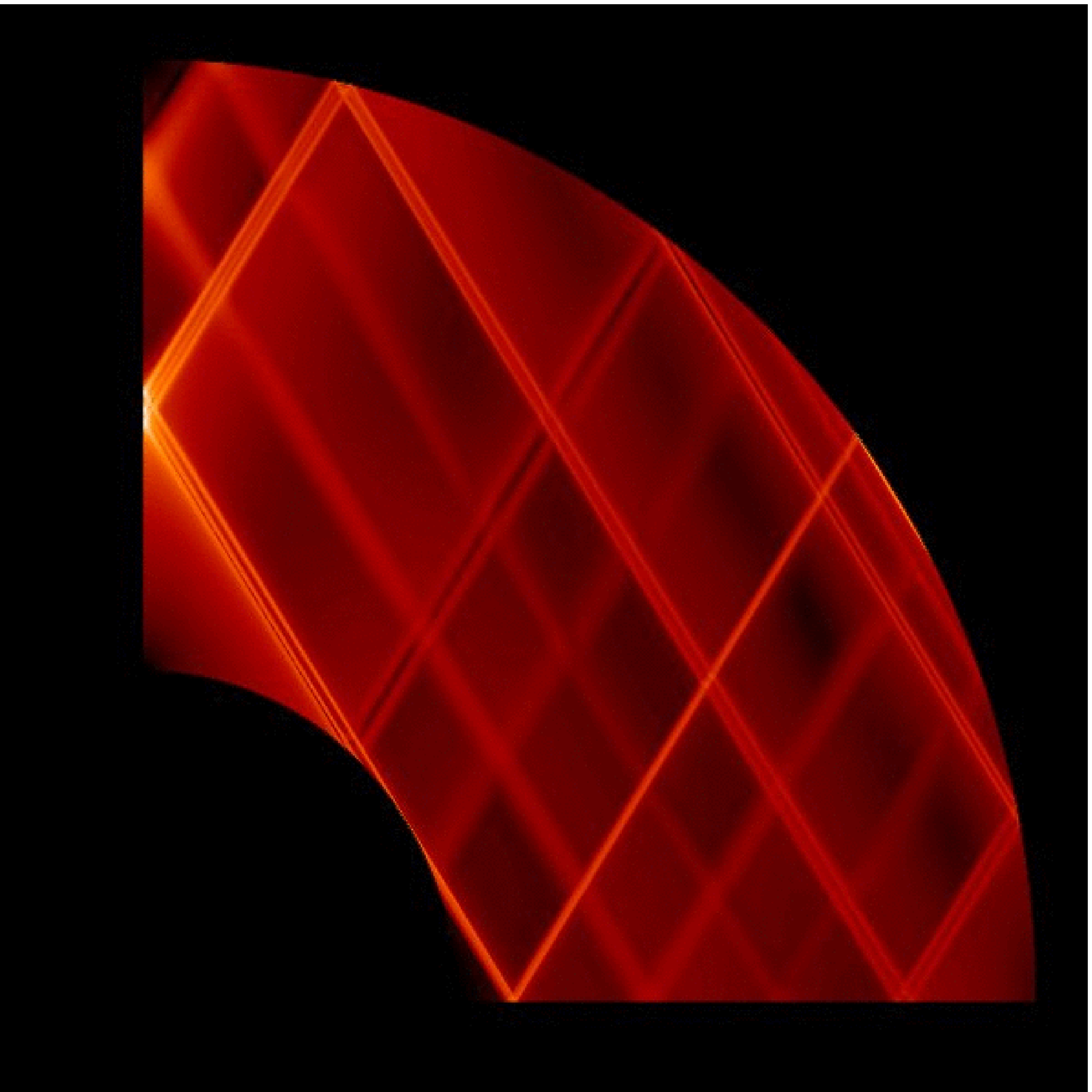}}
\caption[]{Top: Zoom of the dissipation curve for various Ekman numbers
for $\omega\in[0.5,0.6]$. The two vertical solid lines delineate the
frequency range of the attractor shown by the eigenmode below. The dotted
vertical line shows the frequency of the least-damped eigenmode. Below
left: The kinetic energy distribution in a meridional plane of the
least-damped eigenmode ($\omega=0.554838$) associated with the attractor
governing this frequency range.  Below centre: The amplitude of the forced
flow at the frequency of the eigenmode.  Below right: The frequency of
the forcing is now $\omega=0.542$ so that the attractor is closer to
the critical latitude. Note that a second attractor is now visible
and excited by the southern branch of the shear layer emitted at the
critical latitude.  }
\label{zoom_diss555}
\end{figure}

This situation is not unique: it is also the case when we consider the simple
attractor occupying the  frequency band [0.529,0.555].  %[0.528985018136109,0.555369291674672].
As shown in figure~\ref{zoom_diss555}, no resonance occurs
there. Actually, if we compare the least-damped eigenmode associated
with this attractor and the forced flow of the same frequency, which
are shown in the same figure, we clearly see that the attractor is not
excited. There too, we see that the shear layer emitted by the critical
latitude is strongly excited. It propagates towards the attractor.
The dissipation curves computed at various Ekman numbers shows that
for these values of $E$, no asymptotic regime is reached. Actually,
a calculation for a specific frequency shows that an asymptotic regime
exists but at a very low Ekman number ($\infapp 10^{-12}$). We also note
that the Ekman number below which the viscous dissipation is constant
depends on the ``distance" between the attractor and the characteristics
emitted at the critical latitude. We observe that at $\omega=0.542$
the asymptotic regime is almost reached; in this case, the attractor
is much stronger and closer to the critical latitude (as confirmed by
figure~\ref{zoom_diss555}).

Back to the $\omega=0.621$ case, we note that all dissipation curves at
some frequencies above $\omega_0=0.622759$ (which is the upper limit of
the attractor living at $\omega\leq\omega_0$), seem to have converged
to a limit. We think that the asymptotic limit has not actually been
reached there, and that undulations of the dissipation curve will appear,
as they do at higher frequencies, in response to the rapid variations
of the Lyapunov exponent.

To conclude this point, the foregoing examples show that the asymptotic
regime, where dissipation is independent of viscosity, is reached when
the Ekman number is low enough so that the shear layer emitted at
the critical latitude on the inner boundary is ``feeding"
an attractor. In this case the widening and softening of the shear layer,
due to viscous diffusion, is stopped. The shear layers are then in a
regime similar to what has been described by \cite{O05} in two
dimensions.

\subsection{Resonances}

The two-dimensional case gave a nice illustration of resonances which
correspond to frequencies where the Lyapunov exponent vanishes. The
foregoing discussion has shown that the three-dimensional situation is
not so simple. Actually, we could not identify a single resonance that
matches the frequency corresponding to the vanishing Lyapunov exponent
of a well-determined attractor!

\begin{figure}
\centerline{
\includegraphics[width=\linewidth,clip=true]{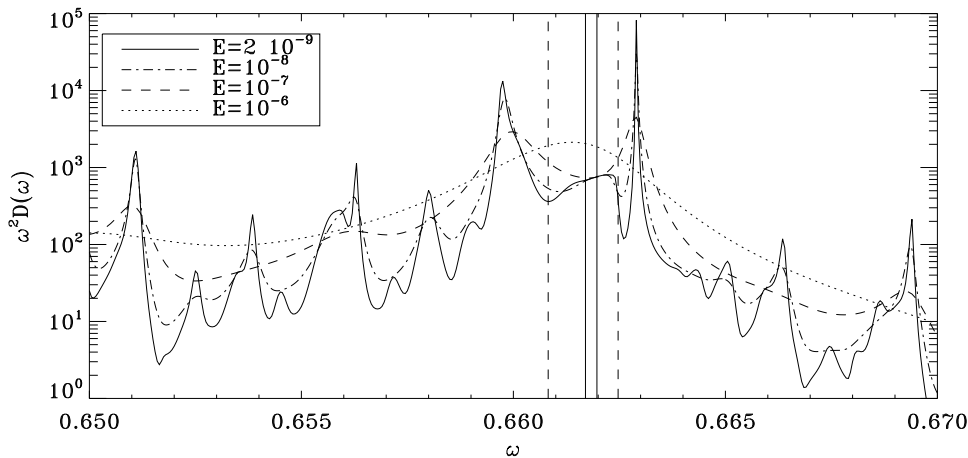}}
\caption[]{Zoom of the dissipation in the spherical shell in a given
frequency range; the four vertical straight lines show the range of existence of two
short-period attractors.}
\label{zoom_diss660}
\end{figure}

Nonetheless, dissipation curves show many peaks indicating some
privileged frequencies. The formation of these peaks as viscosity
decreases is not quite standard.  Indeed, if we compare the dissipation
curves in the interval [0.5,0.6] (fig.~\ref{zoom_diss555}) and the one
in the interval [0.65,0.67],(fig.~\ref{zoom_diss660}) there is a common
feature. This is the fact that ``resonant peaks'' are more and more
numerous as the viscosity decreases (as expected), however they do not
seem to appear because the resonances are more and more vigorous (it is only
marginally the case), but because the neighbouring frequencies resonate
less and less. Hence, it seems to be that many peaks are not true
resonances, but small intervals of frequencies where the dissipation is
independent  of the viscosity provided it is low enough. The series of
peaks for $\omega\in[0.56,0.58]$ is quite illustrative of this phenomenon.

\begin{figure}
\centering
\includegraphics[width=0.7\linewidth,clip=true]{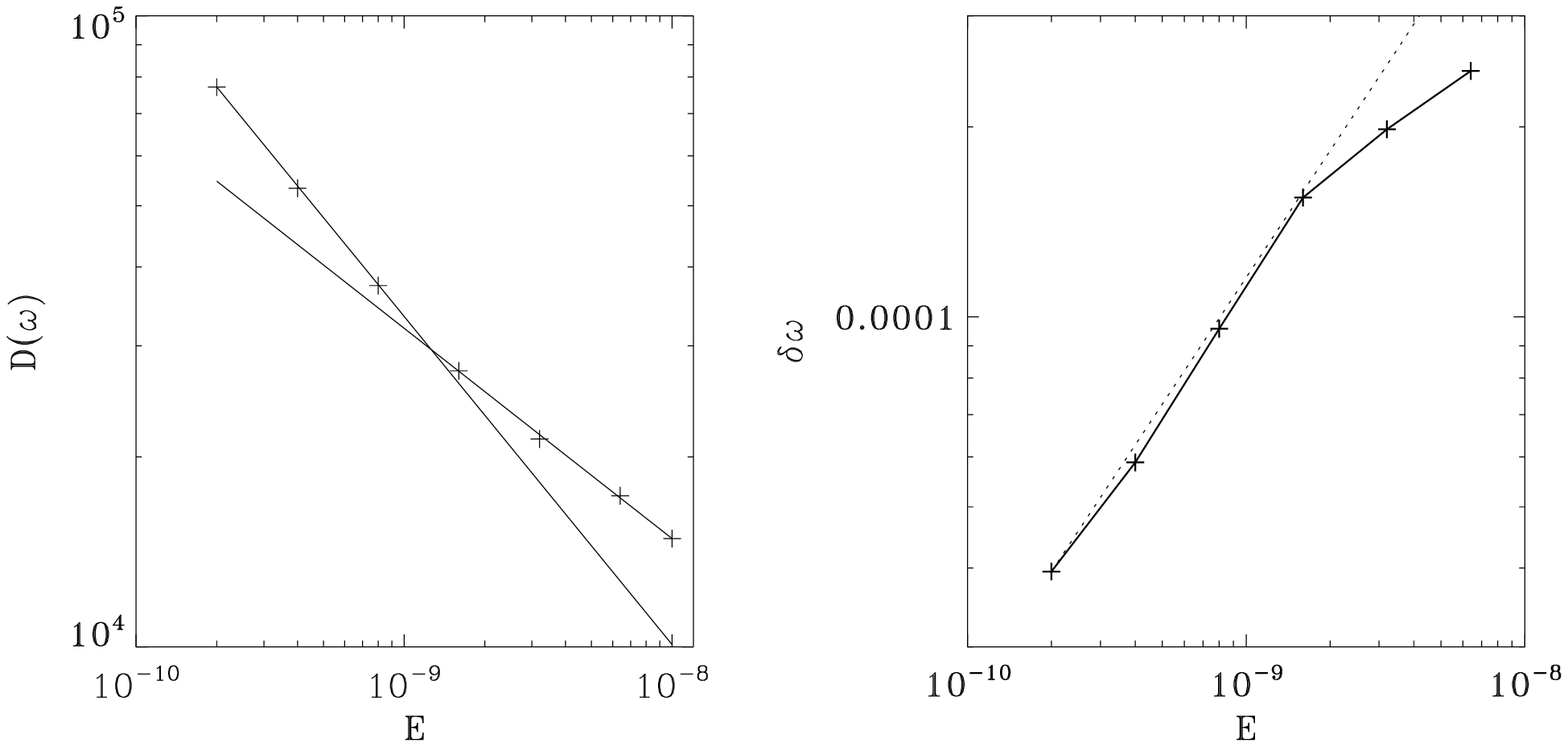}
\includegraphics[width=0.7\linewidth,clip=true]{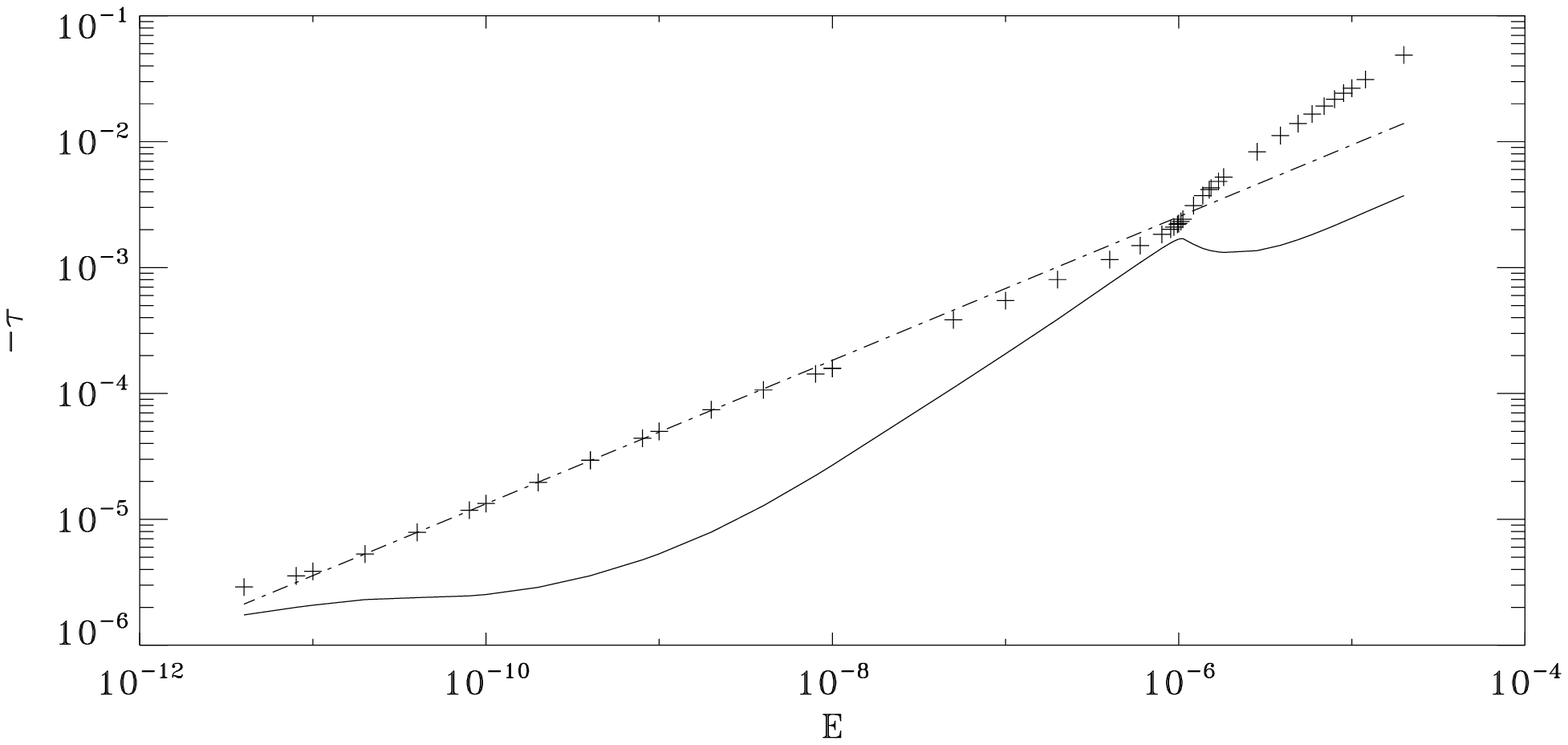}
\caption[]{Top left: The viscous dissipation at the resonance
$\omega=0.6598$ as a function of the Ekman number; the straight lines
shows the power laws E$^{-1/3}$ and E$^{-1/2}$. Top Right: The solid
line and pluses show the width $\delta\omega$ of the resonance compared
to the power law E$^{2/3}$ (dotted line). Below: We show the damping
rate $\tau$ of the two least-damped eigenmodes at $\omega_1=0.6598$
(pluses) and $\omega_2=0.6629$ (solid line);  the power law E$^{0.57}$
is shown by the dash-dotted straight line.} \label{scalings} \end{figure}

\begin{figure}
\centering
\includegraphics[width=0.6\linewidth,clip=true]{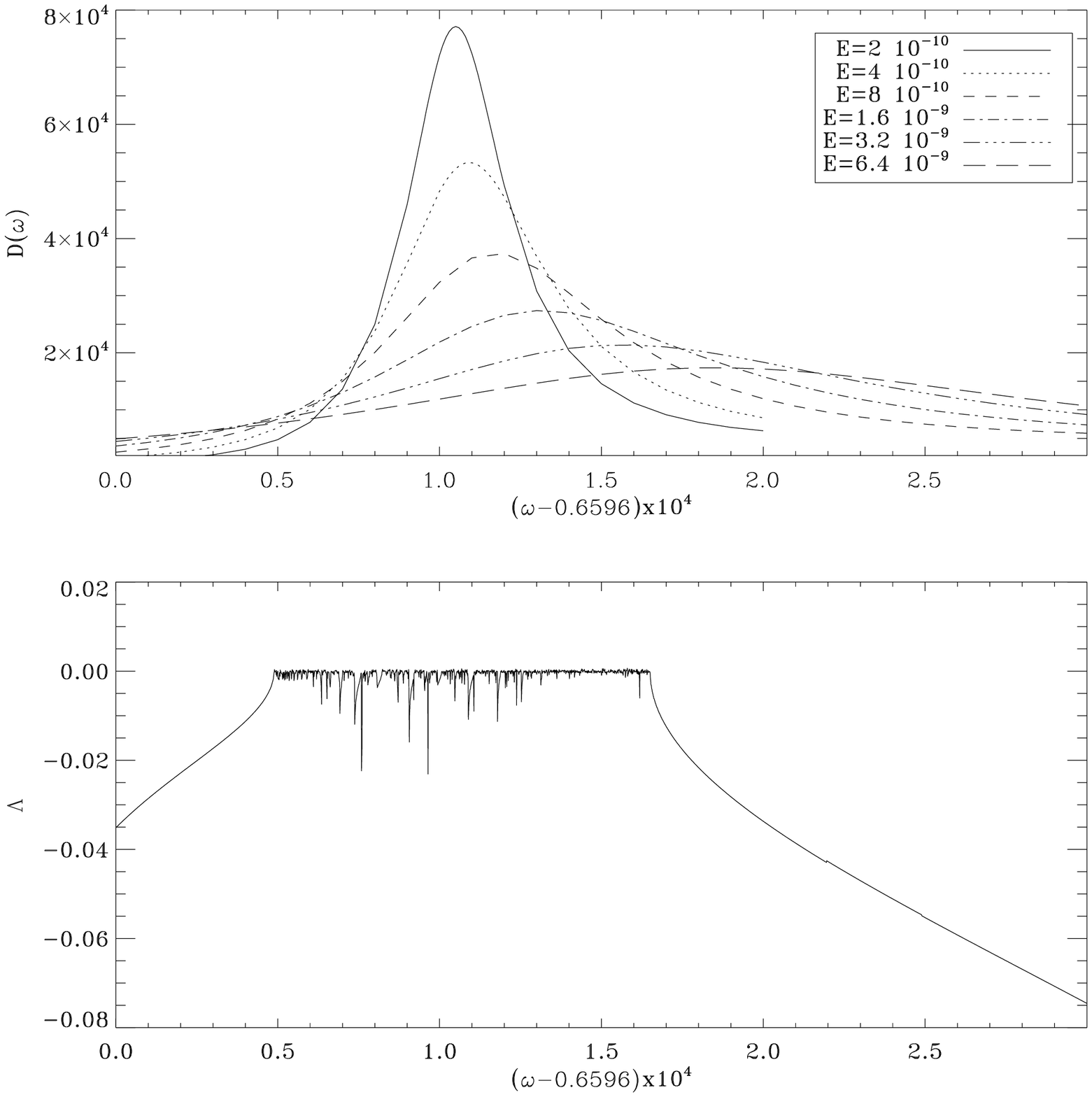}
\caption[]{Top: The viscous dissipation at the resonance near $\omega=0.6596$
as a function of the frequency for various Ekman numbers. Below: The
corresponding Lyapunov exponents in the same frequency interval. Note
the progressive move of the resonance peak towards the frequencies with
small Lyapunov exponents.}
\label{res660}
\end{figure}

However, there are also some peaks which do behave like true
resonances, i.e. their amplitude steadily increases as the Ekman number
decreases. This is for instance the case of the two major peaks of
the spectrum, lying at $\omega_1=0.6598$ and $\omega_2=0.6629$. These
resonances correspond to the least-damped axisymmetric inertial modes that
we actually studied in \cite{RV97}. We therefore extended this previous
study to lower Ekman numbers. In figure~\ref{scalings}, we focus our
attention to the first mode, which seems to follow some asymptotic regime
(as shown, the damping rate of the second mode is clearly not a power
law).  Figure~\ref{scalings} shows that the dissipation, there, grows
like $E^{-1/3}$ when $E\supapp10^{-9}$, but like $E^{-1/2}$ if $E\infapp10^{-9}$.
On the other hand, the width of the resonance narrows as $E^{2/3}$ when
$E\infapp10^{-9}$;
we note that the associated eigenmode has a damping rate decreasing as
$E^{0.57}$.  Applying the simplified model of section~\ref{simple_model}
to this case, we recover the scaling law of the viscous dissipation at
$E\supapp10^{-9}$ if
we assume that the critical latitude singularity emits a shear layer of
width $E^{1/5}$, i.e. corresponding to the latitudinal extension of the
perturbed Ekman layer on the inner sphere. Hence, dissipation should scale
as like $E^{4/5-2\beta}$. Using $\beta=0.57$ we find $D\sim E^{-0.34}$. A
similar matching of the exponents can be obtained if the forced flow
is confined to a neighbourhood of the critical latitude. The volume
responding to the forcing is \od{E^{3/5}} but the velocity gradients are
\od{E^{-2/5}}, thus leading to the same power law for the dissipation.
The simple model of section~\ref{simple_model} cannot be more precise, but
it clearly underlines the crucial role played by the critical latitude.
We refer the reader to the work of \cite{K95} for a detailed analysis of
the boundary layers and shear layers in the vicinity of the critical
latitude in the case of the spin-over mode (i.e the non-axisymmetric, m=1,
at $\omega=0.5$, inertial mode).

Now, the power law change at $E\sim 10^{-9}$ may be understood with
figure~\ref{res660}.  In this figure we show an enlarged view of the
resonance together with the Lyapunov exponents. It is clear that, as the
Ekman number decreases, the resonance shifts to the frequency interval
where this exponent is very small. We note that $E=10^{-9}$ corresponds
to the transition where the resonance leaves the frequency range occupied
by a short-period attractor and enters the frequency interval where only
long-period attractors exist, leading to a stronger resonance. The
new $E^{-1/2}$-regime is likely not asymptotic as well: Inspection of
the Lyapunov curve shows that the size of the frequency intervals,
where a definite Lyapunov exponent exists, is at least less than
$10^{-8}$. This means that an asymptotic regime may be reached at Ekman
numbers less than 10$^{-15}$ (assuming that the resonance keeps narrowing
as $E^{2/3}$). Such numbers, even if theoretically reachable in stars
or planets, are likely unrealistic because any small-scale turbulence
would increase them by many orders of magnitude. Thus, the intermediate
regimes revealed by the $E^{-1/3}$ or $E^{-1/2}$ laws for the dissipation
are likely to be more relevant to astrophysical or geophysical applications,
but more investigations are needed to determine their origine.

\subsection{Influence of the size of the core}

\begin{figure}
\centering
\includegraphics[width=\linewidth,clip=true]{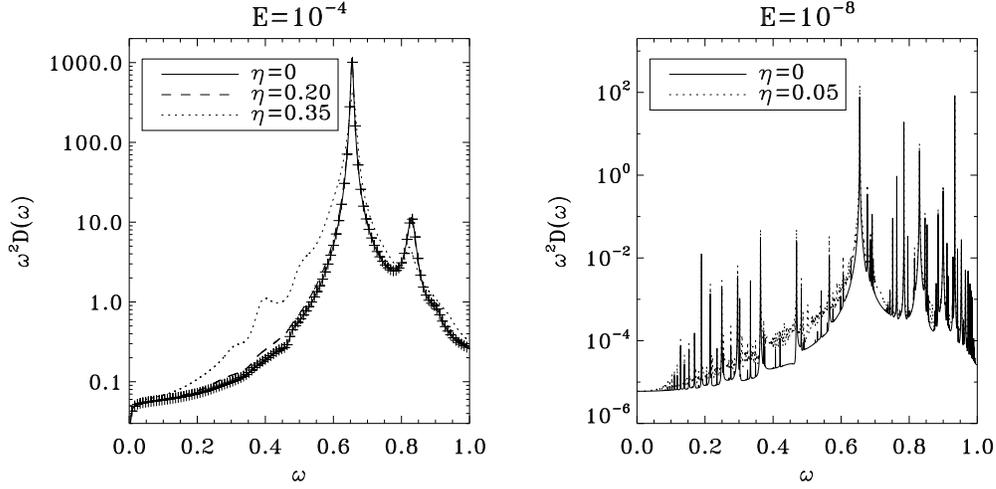}
\caption[]{The resonance curve for various size of the inner core and
two Ekman numbers. Left: Note that the pluses correspond to the case
$\eta=0.1$ and are almost sitting on the curve $\eta=0$; the difference
is always less than two percents. Right: The same scan at a much lower,
but more realistic Ekman number. A small core induces important
differences.}
\label{scan_eta}
\end{figure}

A last question that is often raised is whether the size of the core, when
it is small, influences notably the response curve that would be obtained
if neglecting the core. From the foregoing results and the discussion of
\cite{RGV00}, the answer to this question obviously depends on the Ekman
number. It is indeed expected that for large viscosities, a small inner
core is hardly seen by the fluid motion.  To illustrate further
this point we plot in figure~\ref{scan_eta}, the dissipation curves
for various sizes of the inner core together with the one of the
full sphere. As shown, at a rather large Ekman number $E=10^{-4}$
(unrealistic for planetary or stellar applications), the dissipation
curve of the spherical shell notably differs from the full sphere
case only when $\eta\geq0.2$. On the other hand, with a more realistic
value $E=10^{-8}$, we note that a very small core $\eta=0.05$ already
amplifies the dissipation by an order of magnitude. One also observes
that the frequencies of some resonances are quite similar. This likely
comes from the polynomial nature of the full sphere solutions.  Using a
truncated solution in series of spherical harmonics, \cite{R91} showed
that the frequency of one of the large-scale modes of the full sphere was
shifted by an amount \od{\eta^5}.

In fact, in the inviscid case, the presence of a small core, although
fully perturbating the eigenvalue spectrum of the Poincaré operator,
likely has a less drastic influence on the pseudo-spectrum, which we
partially view through the dissipation curve.

\begin{figure}
\centering
\includegraphics[width=1.0\linewidth,clip=true]{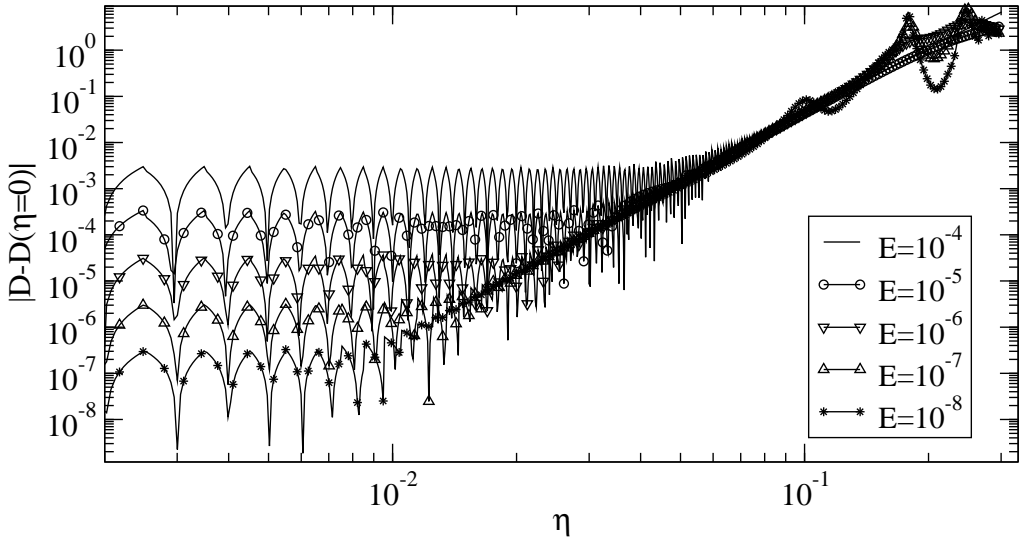}
\caption[]{Variations of the dissipation at a given Ekman number for
varying sizes of the inner core. The frequency of the forcing has been
arbitrarily fixed to $\omega=0.55$.
}
\label{scan_eta_om_fix}
\end{figure}

To be more quantitative and determine the threshold $\eta_c$, for a
given $E$, beyond which the core cannot be ignored, we computed the
dissipations at various $E$, keeping $\omega$ fixed, and varying the
size of the inner core. The curves in figure~\ref{scan_eta_om_fix}
illustrate these variations of the dissipation. They have been computed
for $\omega=0.55$ but the behaviour seems to be generic: We
picked up two other frequencies and found the same trends. However,
for $\omega=\sqrt{3/7}$, which is the broad resonance showing up in
figure~\ref{scan_eta} (right) and which corresponds to a large-scale
mode of the full sphere~\cite[e.g.][]{green69}, we do not find this
behaviour.

From the curves in figure~\ref{scan_eta_om_fix}, we could determine that
for sufficiently small $\eta$'s

\beq D(\eta)=D(\eta=0)+a(\eta)E+b\eta^5
\eeqn{crit_eta}
where $ D(\eta)$ is the viscous dissipation, $b$ is a constant and
$a(\eta)$ is a rapidly varying piecewise bounded linear
function.  The $E$-dependence is expected from a flow which does not
(or little) depend on the viscosity as it is the case for the full sphere
modes. The $\eta^5$-dependence can be, tentatively, explained using the
solution derived in \cite{R91} (see the discussion in appendix C). From
the law \eq{crit_eta} we therefore conclude that the inner core influences
the dissipation when $b\eta^5\geq a_mE$, $a_m$ being the upper bound of
$a(\eta)$, showing that the critical radius
of the inner core is

\beq \eta_c = q E^{1/5}\eeq
where the coefficient $q$ weakly depends on the frequency of the
forcing. We find $q\sim0.4$. This formula shows that only extremely
small cores can be neglected in realistic conditions where $E\infapp
10^{-8}$.

\subsection{Discussion}

The foregoing results show that the three-dimensional flows are
significantly different from their two-dimensional counter part.
The numerical solutions show a kind of exchange of the roles between the
attractors and the critical latitude singularities as one shifts from the
two-dimensional to the three-dimensional problem. In two dimensions
the attractors are clearly dominating the dynamics and control the
periodically forced flows. The shear layer emitted at the critical
latitude is present but not essential. In three dimensions we observe the
opposite: the shear layer emitted at the critical latitude on the inner
sphere plays a crucial role in the fluid's response to the periodic
forcing, while attractors appear when they are ``fed" by this shear
layer.

We have no definite explanation to this observation. We conjecture that it
is a consequence of two modifications in the solutions when one changes
the dimension of the problem. The first may come from the different
Riemann functions of the two- and three-dimensional problems. We recall
that the general solution of an hyperbolic problem may be expressed as

\beqan &&\hspace*{-1.5cm}\lefteqn{\Pi(S)= \demi(\Pi(P)+\Pi(Q))}  \nonumber\\
&&\hspace*{-1cm}+\demi\int_{PQ} R(S,M)\lp
\dup{\Pi}du_+ - \dum{\Pi}du_-\rp + \Pi(M)\lp\dup{R}du_+ -
\dum{R}du_-\rp \eeqan{gen_sol}
where $R(S;M)$ is the Riemann function associated with the Poincaré
operator; integration is done on the data line $PQ$. In 2D this function
is a combination of a Dirac distribution and the identity, while in 3D it
is a Legendre function \cite[see][]{RGV01}.  Hence, in 2D, a singularity
on the data line will remain on the web of characteristics issued from
this point, while in 3D it may generate singular  points outside this
web. Thus, the influence of the critical latitude singularity in the
three-dimensional problem is likely stronger than in the two-dimensional
one.

The second modification brought about by the third dimension is the
boundary of the domain. While in 2D the true fluid domain is a disk
with a core, which is not a simply connected fluid domain, in 3D the meridional
section of the fluid layer is a simply connected domain. The difference comes
from the presence of the rotation axis which is a line of the meridional
plane where the characteristics must reflect, which is not the case
in 2D. We conjecture that this difference makes the eventual existence
of fundamental intervals \cite[introduced by][]{ML95}, more difficult
thus making the excitation of shear layers along the attractors also
more difficult.

\section{Conclusions}

To conclude this work, we may summarise our results in the following
way. First, it turns out that the two-dimensional model, which is
equivalent to the section of a slender (cored) torus, is well understood:

\begin{itemize}
\item The dissipation curve is very spiky as one scans the frequency axis
and each spike corresponds to a resonance. This resonance is associated
with a frequency for which the Lyapunov exponent is zero. We have shown
that in most cases the flow is associated with an attractor whose shear
layer width scales as $E^{1/4}$. In a few cases, when the frequency of the
forcing is of the form $\sin(p\pi/2q)$, where $p$ and $q$ are integers,
the resonance is stronger, but narrower, and corresponds to one of the few eigenmodes
which remain when an inner core is inserted at the centre of a circular domain.

\item In between resonances, which are not dense (but have accumulation
points), the dissipation reaches a constant value at vanishing viscosity,
as predicted by \cite{O05}.

\end{itemize}

The two-dimensional model thus offers a rather neat picture well
constrained by the analytical results. It shows that, contrary to previous
expectations, dissipation is very sensitive to viscosity, essentially
because of resonances but also because, outside resonances, shear
layers associated with different attractors reach their asymptotic regime
at very different Ekman numbers.

In three-dimensions, results show some similarities with the
two-dimensional model but also major differences.  Among the similarities,
we note that the dissipation curve is very spiky too. There also exist
intervals of frequencies where the dissipation does not depend on
viscosity when this quantity is low enough. There are also resonances
where the dissipation seems to increase without bounds when the viscosity
vanishes.

However, there are major differences. First, we observed that the
critical latitude singularity on the inner sphere plays a major role in
the response of the fluid to the forcing. It is systematically emitting
a shear layer. Using the scalings of the Ekman layer at this place,
we could explain the vanishing dissipation with vanishing viscosity
when the frequency of the forcing is associated with a periodic orbit
of characteristics. The latter case also stresses another difference:
whereas in the two-dimensional case these periodic orbits were associated
with strong and narrow resonances, in the three-dimensional case they
are associated with anti-resonances : the dissipation vanishes with the
viscosity (following a $E^{2/5}$-law).

We observed that in the two-dimensional case, least-damped modes are
associated with frequencies where the Lyapunov exponent is zero.  In three
dimensions, such an association also exists but some of the least-damped
modes are also found in association with intervals of frequencies where
the mapping is weakly contracting (because of long-period attractors). Our
numerical results on the forced problem show that only this latter
kind of least-damped modes lead to strong resonances.  The question of
whether these resonances can reach an asymptotic regime remains open;
the fractal nature of the Lyapunov exponents in the frequency region
where they exist may prevent any asymptotic limit.  Clearly, for one of
them, even at $E=10^{-11}$, the asymptotic regime is not reached. The
close examination of this resonance has nevertheless shown that some
intermediate regimes may exist when attractors of very long period cover
the same intervals of frequencies as the resonances. Such regimes are
astrophysically relevant and deserve more investigations.

Unfortunately and unlike in the two-dimensional case, we have
little analytical guide in the three-dimensional case. Tentatively,
we interpreted the striking role of the characteristics emitted at
the critical latitude as a consequence of the nature of the Riemann
function associated with the Poincaré operator and the probable different
fundamental intervals associated with the mapping of characteristics.

Finally, the striking general conclusion is that a periodically forced
flow in a spherical shell is quite different from what can be expected
from a simple response of resonant eigenmodes. Previous work has shown
that eigenmodes were often featured by attractors of characteristics, but
those modes turned out to be of little interest for the forced problem.
Interestingly enough, the dissipation curve reveals some parts of the
pseudospectrum of the Poincaré operator, thus offering another
way to investigate the properties of this operator.

The solutions that we computed also showed that the critical latitude
singularity is a determinant feature of the periodically forced flows.
More work is now needed to fully understand the interplay of the
singularities generated by both the critical latitude and the attractors
of characteristics. We note that \cite{GL09}, while examining the tidal
response of a Jovian planet, were also led to the conclusion that the
critical latitude at the interface between a solid core and a fluid envelope
plays a crucial role in the dissipation of kinetic energy. In this same
planetary context, we showed that planets with a very small core could be
assimilated to a full sphere only when the relative radius of the inner
core, $\eta$, is less than $0.4E^{1/5}$. Beyond this, usually very
small, value the dissipation grows rapidly with the core size as
$\eta^5$.

To conclude on the astrophysical problem, which motivated this study,
our results stress the importance of viscosity. They unfortunately remove
the possibility, envisaged by \cite{OL04}, that the synchronisation time
scale of binary stars be independent of the viscosity, which is usually
not a well known quantity.

\begin{acknowledgements}
We are very pleased to acknowledge fruitful discussions with Gordon
Ogilvie and Serge Gratton. We are also grateful to Keke Zhang and the referees for
their detailed review of the first version of this work.  The numerical
calculations have been carried out on the NEC SX8 of the `Institut du
Développement et des Ressources en Informatique Scientifique' (IDRIS)
and on the CalMip machine of the `Centre Interuniversitaire de Calcul
de Toulouse' (CICT) which are both gratefully acknowledged.
\end{acknowledgements} 

\bibliography{../../../biblio/bibnew}

\appendix
\section{Viscous dissipation in the slender torus}

\subsection{Equivalence of coordinate systems}

Using non-dimensional variables, viscous dissipation is

\[ D  = \frac{E}{2}\intvol|c|^2 dV \]
where $|c|$ symbolises the norm of the shear tensor. In usual cylindrical
coordinates, this norm reads

\[|c|^2 = c_{ij}c_{ij}^* = |c_{ss}|^2+|c_{\varphi\varphi}|^2+|c_{zz}|^2 +
2(|c_{s\varphi}|^2+|c_{sz}|^2+|c_{\varphi z}|^2)\]
with

\[ c_{ss} = 2\ds{v_s}, \quad c_{\varphi\varphi}=2\lp\dsvarphi{v_\varphi} +
\frac{v_s}{s}\rp, \quad c_{zz} = 2\dz{v_z}\]
\[ c_{s\varphi} = \dsvarphi{v_s} + \ds{v_\varphi} -
\frac{v_\varphi}{s}, \qquad c_{sz} = \ds{v_z} + \dz{v_s}, \qquad
c_{\varphi z} = \dz{v_\varphi} + \dsvarphi{v_z}\]
Considering axisymmetric solutions at $s\gg1$, the shear tensor components
simplify into

\[ c_{ss} = 2\ds{v_s}, \qquad c_{\varphi\varphi}=0, \quad c_{zz} =
2\dz{v_z}\]
\[ c_{s\varphi} = \ds{v_\varphi}, \qquad c_{sz} = \ds{v_z} + \dz{v_s},
\qquad
c_{\varphi z} = \dz{v_\varphi} \]
Using the meridional stream function $\psi$, we may further write

\[c_{zz} = 2\frac{\partial^2\psi}{\partial s\partial z} = - c_{ss} \]

\[c_{sz}= \dds{\psi} - \ddz{\psi} \]

\[c_{s\varphi} = \ds{u},\qquad c_{\varphi z} = \dz{u} \]
So that the volumic dissipation is :

\[|c|^2 = 2(|c_{ss}|^2+|c_{s\varphi}|^2+|c_{sz}|^2+|c_{\varphi z}|^2)\]

Now a direct calculation shows that:

\[c_{zz} =
\sin2\phi\lp\psi_{\rho\rho}-\psi_\rho/\rho-\psi_{\phi\phi}/\rho^2\rp
+ \frac{2\cos2\phi}{\rho}\lp\psi_{\rho\phi} -
\psi_\phi/\rho\rp = -c_{ss} \]
and

\[ c_{sz} = \cos2\phi(\psi_{\rho\rho}-\psi_\rho/\rho-
\psi_{\phi\phi}/\rho^2) - \frac{2\sin2\phi}{\rho}\lp\psi_{\rho\phi}
-\psi_\phi/\rho\rp\]
where we recognise the expressions of $c_{\rho\rho}$ and $c_{\rho\phi}$ of \eq{}. Thus,

\[c_{zz} = -c_{ss} = \sin2\phi c_{\rho\phi} +\cos2\phi c_{\rho\rho} \]
\[  c_{sz} = \cos2\phi c_{\rho\phi} - \sin2\phi  c_{\rho\rho} \]
In the same way, we also find that

\[ c_{s\phi}=\cphi c_{\rho \zeta} -\sphi c_{\phi \zeta} \andet c_{\phi z} =
\sphi c_{\rho\zeta}+\cphi c_{\phi\zeta}\]
which leads to 

\[ |c|^2 = 2
(|c_{\rho\rho}|^2+|c_{\rho\zeta}|^2+|c_{\phi\zeta}|^2+|c_{\rho\phi}|^2)\]
as expected.

\subsection{Dissipation as function of the Fourier components}

We first note that

\begin{eqnarray*}
\int_0^{2\pi}|c_{\rho\zeta}|^2+|c_{\theta\zeta}|^2d\theta &=&
 2\pi \sum_n|V'_n|^2 + \frac{n^2}{\rho^2}|V_n|^2 \\
\int_0^{2\pi}|c_{\rho\rho}|^2 d\theta &=& 2\pi \sum_n
\left|\frac{2n}{\rho}\lp\psi'_n - \frac{\psi_n}{\rho}\rp\right|^2 \\
\int_0^{2\pi}|c_{\rho\theta}|^2 d\theta &=& 2\pi \sum_n \left|\psi''_n -
\frac{\psi'_n}{\rho}+\frac{n^2\psi_n}{\rho^2}\right|^2
\end{eqnarray*}
Since the total dissipation is

\[
D=E\intvol(|c_{\rho\rho}|^2+|c_{\rho\zeta}|^2+|c_{\theta\zeta}|^2+|c_{\rho\theta}|^2)dV,
\]
it finally expresses as

\[ D =2\pi E\int_\eta^1 \left\{ \sum_n|V'_n|^2 +
\frac{n^2}{\rho^2}|V_n|^2 +\left|\frac{2n}{\rho}\lp\psi'_n -
\frac{\psi_n}{\rho}\rp\right|^2 +\left|\psi''_n -
\frac{\psi'_n}{\rho}+\frac{n^2\psi_n}{\rho^2}\right|^2\right\}\rho
d\rho\]
If the solutions are equatorially symmetric $V_{-n}=V_n$ and $\psi_{-n}=-\psi_n$, so that

\[ D =4\pi E\int_\eta^1\frac{}{} \left\{
\demi|V'_0|^2+\sum_{n>0}\frac{}{}|V'_n|^2 +
\frac{n^2}{\rho^2}|V_n|^2 +\left|\frac{2n}{\rho}\lp\psi'_n -
\frac{\psi_n}{\rho}\rp\right|^2 +\left|\psi''_n -
\frac{\psi'_n}{\rho}+\frac{n^2\psi_n}{\rho^2}\right|^2\right\}\rho
d\rho\]

\section{Influence of boundary conditions}

\cite{FH98} studied the influence of boundary conditions (no-slip or
stress-free) on the inertial modes of a spherical shell. They found that
this influence was very small. In the forced flow that we study here,
this is also the case. The contribution to dissipation of Ekman layers
that appear if no-slip boundary conditions are used may be estimated as
follows: In the case of the asymptotic régime where the fluid flow is
following an attractor, the velocity scales as $E^{-1/3}$ \cite[][]{O05}
and the boundary layer velocity gradients $\partial u/\partial x$ are
\od{E^{-1/3}/E^{1/2}} = \od{E^{-5/6}}. However, the volume occupied
by the boundary layers is \od{E^{5/6}}, thus, the resulting dissipation
scales like $E^{1/6}$ which is very small compared to the contribution of
the internal shear layers, which is unity. In the case of the resonances
of the 2D-slender torus, which are associated with shear layers of width
scaling in $E^{1/4}$, the contribution of Ekman layers is even weaker,
namely \od{E^{1/4}}. It is only in the case of the regular modes of
the torus, that a strong effect is noticed, as expected. However these
modes do not exist in the more realistic set-up of a spherical shell: we
obtained singular solution (the ``anti-resonances" see \S~\ref{antir}).
In this case we find that the dissipation scales like $E^{2/5}$, which
seems to be also the case when a no-slip boundary condition is used on
the inner core.

\section{The asymptotic law for cores of vanishing sizes}

We may recover the dependence in $\eta^5$ of the viscous dissipation when
$\eta\ll1$ by considering the solutions found by \cite{R91}. Indeed, in
this paper it was shown that an oscillatory flow in a rotating background
with spherical symmetry could be described by two kinds of solutions when
these are decomposed onto the spherical harmonic basis. These solutions
(in fact the radial functions of each spherical harmonic component) are
either of Bessel or polynomial type. The Bessel type solutions describe
the boundary layer regions, while the polynomials describe the flow in
the remaining volume.

Considering the inviscid case, we shall focus on the polynomial
solutions. \cite{R91} has shown that these solutions also split into
two categories: one is regular at the origin, the other is regular
at infinity. The first one is easily related to the inertial modes
of the full sphere which have been found by \cite{Bryan1888} \cite[see
also][]{green69,R97}. They are exact solutions of the inviscid problem at
eigenfrequencies. Below, we shall refer to these solutions as class-one
solutions.

The second class of solutions are those with radial functions with
polynomials in $1/r$ that are regular at infinity. However, unlike those
regular at the origin, these solutions are exact only when the momentum
equation is expanded on a finite number of spherical harmonics, namely
when the series is artificially truncated at a given maximum order, just
like in a numerical calculation. This is expected since the solutions
regular at infinity are singular at the critical latitude on an inner
bounding sphere \cite[see][]{RGV01}. Below, we refer to these solutions
as class-two solutions.

Now if we think to the solutions in their spherical harmonic
decomposition, we note that the radial velocity of the class-one solution
is dominated near the centre by its $u^2Y_2$ component (we restrict
the discussion to the axisymmetric and equatorially symmetric flows).
This is because $\ul\sim r^{\ell-1}$ as $r\tv0$. When a small inner core
is introduced in the inviscid problem, we need to add solutions that are
regular at infinity so that the inner boundary condition $u_r(\eta)=0$
is met. This is the usual procedure in regular elliptic problems like
the Poisson problem for instance. However, in the Poincaré problem, this
cannot be done because the class-two solutions are singular. However,
if we restrict our problem to a finite number of spherical harmonics
(which is a way of regularization!), we can use this kind of solutions.

As shown in \cite{R91}, the form of class-2 solutions for $u^2(r)$ is in $r^{-4}$. If we
write

\[ \vu = \vu_0 + \alpha_2\vu^2_\infty+\alpha_4\vu^4_\infty+\cdots\]
where $\vu_0$ is the class-1 solution and $\vu^J_\infty$ are the class-2
solution of order $J$ (their lowest order is $u^J(r)=r^{-J}$). Imposing
that $u_r(\eta)=0$ for $\eta\ll1$, we easily find that $\alpha_2$ is
\od{\eta^5} and that more generally, $\alpha_{2p}$ is \od{\eta^{2p+3}}.

Hence, to leading order, we may write

\beq \vu = \vu_0 + \eta^5\vu_\eta\eeqn{expan_eta}
Although the truncated solutions are not exact, we know that viscosity
will, in the end, also truncate the spherical harmonics series. Computing
the viscous dissipation from \eq{expan_eta} shows that

\[ D(\eta) = D(\eta=0)+\eta^5D_1\]
where the $D_1$-term might depend on the Ekman number. Although quite
close to the searched expression~\eq{crit_eta}, this equation misses the
$a(\eta)E$ term and the fact that $D_1$ seems to be independent of $E$.

We note that the critical latitude boundary layer on the inner sphere
could induce a velocity field which is \od{\eta/E^{1/5}}. This scaling
also suggests that a transition occurs when $\eta\sim E^{1/5}$. This
latter remark shows that the complete explanation is likely involved and a
more detailed analysis is necessary to fully understand the intricacy
of the limits $\eta\tv0$ and $E\tv0$.

\end{document}

%% file: com.tex
\def\bbbn{{\rm I\!N}} %natuerliche Zahlen
\newcommand{\andet}{ {\qquad {\rm and}\qquad}}

\newcommand{\tv}{ \rightarrow}

\newcommand{\BVF}{Brunt-V\"ais\"al\"a frequency\ }

\newcommand{\beq}{\begin{equation}}
\newcommand{\beqa}{\begin{eqnarray*}}
\newcommand{\beqan}{\begin{eqnarray}}
\newcommand{\greq}{\begin{equation}\left. \begin{array}{l}}
\newcommand{\egreq}{\end{array}\right\} \end{equation}}
\newcommand{\nngreq}{\[\left\{ \begin{array}{l}}
\newcommand{\nnegreq}{\end{array}\right. \]}
\newcommand{\egreqn}[1]{\end{array}\right\} \label{#1}\end{equation}}
\newcommand{\eeq}{\end{equation}} 
\newcommand{\eeqn}[1]{\label{#1}\end{equation}} 
\newcommand{\eeqa}{\end{eqnarray*}}
\newcommand{\eeqan}[1]{\label{#1}\end{eqnarray}}

\newcommand{\lp}{ \left(}
\newcommand{\rp}{ \right)}

\newcommand{\la}{ \left\{ }
\newcommand{\ra}{\right\} }

\newcommand{\Deltal}{\Delta_\ell}
\newcommand{\ul}{u^\ell}

\newcommand{\wl}{w^\ell}
\newcommand{\ulmm}{u^{\ell-1}}
\newcommand{\wlmm}{w^{\ell-1}}
\newcommand{\ulp}{u^{\ell+1}}

\newcommand{\ulm}{u^\ell_m}
\newcommand{\vlm}{v^\ell_m}
\newcommand{\wlm}{w^\ell_m}

\newcommand{\YL}{ Y^m_\ell }

\newcommand{\RL}{ \vec{R}^m_\ell }
\newcommand{\SL}{ \vec{S}^m_\ell }
\newcommand{\TL}{ \vec{T}^m_\ell }

\newcommand{\na}{ \vec{\nabla} }

\newcommand{\intvol}{ \int_{(V)}\! }
\newcommand{\cth}{ \cos\theta }
\newcommand{\sth}{ \sin\theta }
\newcommand{\cphi}{ \cos\phi }
\newcommand{\sphi}{ \sin\phi }

\newcommand{\vu}{\vec{u}}

\newcommand{\er}{\vec{e}_r}

\newcommand{\ephi}{\vec{e}_\phi}

\newcommand{\ez}{\vec{e}_z}

\newcommand{\vf}{\vec{f}}

\newcommand{\vn}{\vec{n}}

\newcommand{\vR}{\vec{R}}

\newcommand{\vT}{\vec{T}}
\newcommand{\vv}{\vec{v}}
\newcommand{\vV}{\vec{V}}

\newcommand{\vX}{\vec{X}}

\newcommand{\vO}{\vec{\Omega}}

\newcommand{\calA}{{\cal A}}

\newcommand{\calL}{{\cal L}}
\newcommand{\demi}{\frac{1}{2}}

\newcommand{\od}[1]{\mbox{${\cal O}(#1)$}}

\newcommand{\dr}[1]{\frac{\partial  #1}{\partial r}}
\newcommand{\drho}[1]{\frac{\partial  #1}{\partial\rho}}
\newcommand{\drhorho}[1]{\frac{\partial^2  #1}{\partial\rho^2}}
\newcommand{\ds}[1]{\frac{\partial  #1}{\partial s}}

\newcommand{\ddnr}[1]{\frac{d^2  #1}{dr^2}}

\newcommand{\dup}[1]{\frac{\partial  #1}{\partial u_+}}
\newcommand{\dum}[1]{\frac{\partial  #1}{\partial u_-}}

\newcommand{\dz}[1]{\frac{\partial  #1}{\partial z}}

\newcommand{\dtheta}[1]{\frac{\partial  #1}{\partial \theta}}

\newcommand{\dsvarphi}[1]{\frac{1}{s}\frac{\partial  #1}{\partial\varphi}}
\newcommand{\ddr}[1]{\frac{\partial^2  #1}{\partial r^2}}

\newcommand{\ddrho}[1]{\frac{\partial^2  #1}{\partial\rho^2}}

\newcommand{\drhophi}[1]{\frac{1}{\rho}\frac{\partial #1}{\partial\phi}}
\newcommand{\drhodphi}[1]{\frac{\partial^2 #1}{\partial\rho\partial\phi}}

\newcommand{\dds}[1]{\frac{\partial^2  #1}{\partial s^2}}
\newcommand{\ddz}[1]{\frac{\partial^2  #1}{\partial z^2}}

\newcommand{\dphi}[1]{\frac{\partial  #1}{\partial \phi}}

\newcommand{\ddphi}[1]{\frac{\partial^2  #1}{\partial \phi^2}}

\newcommand{\Div}{\vec{\nabla}\cdot}

\newcommand{\eq}[1]{(\ref{#1})}

\newcommand{\infapp}{\raisebox{-.7ex}{$\stackrel{<}{\sim}$}}
\newcommand{\supapp}{\raisebox{-.7ex}{$\stackrel{>}{\sim}$}}
\def\og{\leavemode\raise.3ex\hbox{$\scriptscriptstyle\langle\!\langle$}}
\def\fg{\leavemode\raise.3ex\hbox{$\scriptscriptstyle\rangle\!\rangle$}}